\RequirePackage[final]{graphicx}
\documentclass[aps, superscriptaddress, prx, twocolumn]{revtex4-2}
\usepackage[utf8]{inputenc}
\usepackage{amsmath}
\usepackage{amssymb}
\usepackage{amsfonts}
\usepackage{mathtools}
\usepackage{fixme}
\usepackage{color}
\usepackage{braket}
\usepackage{dsfont}
\usepackage{hyperref}
\usepackage{url}

\usepackage{tikz}
\usetikzlibrary{calc}

\usetikzlibrary{positioning,arrows}
\usetikzlibrary{decorations}
\usetikzlibrary{decorations.pathmorphing,positioning}
\usetikzlibrary{decorations.markings}

\usepackage{tgtermes} 

\tikzset{square left brace/.style={ncbar=0.1cm}}
\tikzset{square right brace/.style={ncbar=-0.1cm}}

\definecolor{myred}{RGB}{214,26,70}
\definecolor{myreddark}{RGB}{76,8,38}
\definecolor{myblue}{RGB}{35,106,185}
\definecolor{mybluedark}{RGB}{19,56,99}
\definecolor{mybluebright}{RGB}{225,236,249}

\def\te{{\rm e}}

\def\nn{\nonumber}

\def\Im{{ \rm Im }}

\def\AF{{ \rm AF }}
\def\Ham{{ \hat{H} }}

\begin{document}
\title{Exact dynamics of two holes in two-leg antiferromagnetic ladders}
\date{\today}

\author{K. \ Knakkergaard \ Nielsen}
\affiliation{Max-Planck Institute for Quantum Optics, Hans-Kopfermann-Str. 1, D-85748 Garching, Germany}

\begin{abstract}
We study the motion of holes in a mixed-dimensional setup of an antiferromagnetic ladder, featuring nearest neighbor hopping $t$ along the ladders and Ising-type spin interactions along, $J_\parallel$, and across, $J_\perp$, the ladder. We determine exact solutions for the low-energy one- and two-hole eigenstates. The presence of the trans-leg spin coupling, $J_\perp$, leads to a linear confining potential between the holes. As a result, holes on separate legs feature a super-linear binding energy scaling as $(J_\perp / t)^{2/3}$ in the strongly correlated regime of $J_\perp,J_\parallel \leq t$. This behavior is linked to an emergent length scale $\lambda \propto (t/J_\perp)^{1/3}$, stemming from the linear confining potential, and which describes how the size of the two-hole molecular state diverges for $J_\perp,J_\parallel \ll t$. On the contrary, holes on the same leg unbind at sufficiently low spin couplings. This is a consequence of the altered short-range boundary condition for holes on the same leg, yielding an effective Pauli repulsion between them, limiting their kinetic energy and making binding unfavorable. Finally, we determine the exact nonequilibrium quench dynamics following the sudden immersion of initially localized nearest neigbhor holes. The dynamics is characterized by a crossover from an initial ballistic quantum walk to an aperiodic oscillatory motion around a finite average distance between the holes due to the confining potential between them. In the strongly correlated regime of low spin couplings, $J_\perp, J_\parallel \leq t$, we find this asymptotic distance to diverge as $t / J_\perp$, showing a much stronger scaling than the eigenstates. The predicted results should be amenable to state-of-the-art quantum simulation experiments using currently implemented experimental techniques. 
\end{abstract}

\maketitle

\section{Introduction}
Quantum simulation experiments have matured to the level, where they push our understanding of many-body quantum dynamics and inspire new approximate theoretical tools \cite{Chen2021,Sinha2022,Grusdt2018_2,Blomquist2020,Nielsen2022_2} that allow us to explore the complex spatial structures arising in e.g. Fermi-Hubbard systems \cite{2010Esslinger,Boll2016,Cheuk2016b,Mazurenko2017,Hilker2017,Brown2017,Chiu2018,Brown2019,Koepsell2019,Chiu2019,Brown2020a,Vijayan2020,Hartke2020,Brown2020b,Koepsell2021,Ji2021,Gall2021}. A major driver for this line of research is to better understand the microscopic origins of high-temperature superconductivity \cite{highTc}, which basic phenomenology may be explained by the interaction and ensuing pairing of dopants in Fermi-Hubbard systems \cite{Emery1987,Schrieffer1988,Dagotto1994}. Recent experiments \cite{Hirthe2023} have for the first time successfully demonstrated that cold-atom simulators can achieve and probe the formation of such pairs in a particular kind of spin ladders. Whereas these experiments were still limited to rather small system sizes, they have shown a great promise in how we can understand these mechanisms from the bottom-up perspective, and the approximate theoretical description of this situation \cite{Bohrdt2022} suggests that the system supports a strong binding of the dopants, in contrast to the usual scenario in two-dimensional square lattice geometries \cite{Eder1992,Boninsegni1993,Poilblanc1994,Chernyshev1998,Hamer1998,Vidmar2013,Mezzacapo2016}. Importantly, spin ladders also arise in compound materials supporting unconventional superconductivity \cite{Uehara1996,Dagotto1992}. While these compounds are mainly probed in scattering experiments \cite{Eccleston1998,Notbohm2007,Lake2010,Schlappa2009,Schlappa2018,Kumar2019}, cold-atom simulators give direct access to spatial correlations and nonequilibrium dynamics \cite{Hilker2017,Vijayan2020,Koepsell2021}.

%%%%%%%%%%%%%%%%%%%%%%%%%%%%%%%%%%%%%%%%%%%%%%%%%%%%%%%%%%%%%%%%%%% 
\begin{figure}[t!]
\begin{center}
\includegraphics[width=0.9\columnwidth]{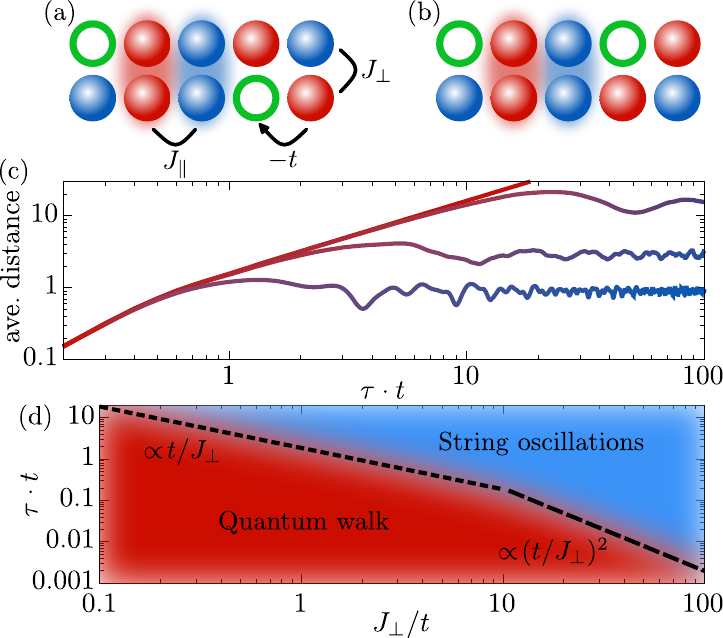}
\end{center}\vspace{-0.5cm}
\caption{Mixed-dimensional $t$--$J$ model featuring spin-$1/2$ particles on a two-leg ladder geometry with two holes on separate legs (a) or on the same leg (b). The spins can hop to nearest neighbor vacant sites along the ladder with amplitude $-t$, and have nearest neighbor Ising interactions $J_\parallel, J_\perp$ along and across the ladder, respectively. (c) Average distance between two holes versus time $\tau$, which initially sit at nearest neighbor sites. At short times, the holes blow apart ballistically as described by a quantum walk [red line], after which they oscillate around a well-defined long-time average. This is shown for $J_\perp/t = 0.2,1,3$ from top to bottom. (d) Corresponding dynamical regimes: quantum walk at short times [red region] and confining string oscillations [blue region] on long timescales. The crossover scales as $(t/J_\perp)^2$ and $t/J_\perp$ in the weak [$J_\perp \gg t$, dashed line] and strong [$J_\perp \ll t$, long-dashed line] correlation regime. The lines in (c) are colored to match the dynamical regimes in (d).}
\label{fig.introfig} 
\vspace{-0.25cm}
\end{figure} 
%%%%%%%%%%%%%%%%%%%%%%%%%%%%%%%%%%%%%%%%%%%%%%%%%%%%%%%%%%%%%%%%%%%

Inspired by this development, we analyze a situation, in which we may gain \emph{exact} results for the binding and nonequilibrium dynamics of dopants in a mixed-dimensional Fermi-Hubbard system [Figs. \ref{fig.introfig}(a)-\ref{fig.introfig}(b)]. The main theoretical difficulty in previous studies \cite{Bohrdt2022} has been to fully describe isotropic spin couplings, coming with the complication of an underlying order-disorder phase transition as the trans-leg spin coupling is increased \cite{hida1992,sandvik1994,scalettar1994,sandvik1995,chubukov1995,Gall2021}. However, by restricting the spin interactions to the Ising type, the underlying spin lattice always supports a perfectly Ne{\'e}l-ordered ground state. Based on this simplification, we find that we can describe the low-energy single- and two-hole eigenstates exactly in this case, whether they be on the same or separate legs [Figs. \ref{fig.introfig}(a)-\ref{fig.introfig}(b)]. Furthermore, using the precise insights into the two-hole eigenstates, we calculate the exact quench dynamics following the sudden creation of two holes as nearest neighbors. Here, Figs. \ref{fig.introfig}(c)-\ref{fig.introfig}(d) show the result of holes on separate legs. We find that the dynamics can be divided into two characteristic regimes. First, the holes perform independent quantum walks, meaning that they blow apart ballistically. Second, as the holes diverge from each other, a confining string of overturned spins forms between them. Eventually, the holes are slowed down by this confinement and aperiodic oscillations in the strings, and thereby in the inter-hole distance, take place around a well-defined long-time average.

A major challenge in previous experiments with doped Fermi-Hubbard systems \cite{Koepsell2019,Ji2021} has been to reach the strongly correlated regime, which is interesting both from the perspective of the physics of the cuprate materials supporting high-temperature superconductivity \cite{highTc}, and for understanding many-body phenomena outside the realm of perturbation theory. We note that this system is a natural experimental candidate for that, because the effective coupling strength between the holes is $4t / J_\perp$. Consequently, the crossover timescale from the quantum walk to the string oscillation behavior in Fig. \ref{fig.introfig}(d) changes from a perturbative $(t/J_\perp)^2$ scaling to a strongly correlated scaling of $t/J_\perp$ already for $J_\perp \lesssim 4t$. Importantly, the crossover time is still quite moderate in terms of hopping times $1 / t$, and remains below $3/t$ for $J_\perp > t$, which should make it possible to experimentally observe the departure from the quantum walk. While the mixed-dimensional property of this model has already been achieved experimentally \cite{Hirthe2023}, the ability to tune the spin interactions to the Ising limit can be facilitated by Rydberg-dressed atoms \cite{Glaetzle2015,Bijnen2015,Zeiher2016,Zeiher2017,Borish2020,Sanchez2021}, polar molecules \cite{Gorshkov2011_2}, and trapped ions \cite{Britton2012}. This setup is, therefore, within reach for modern quantum simulation experiments. \\

The paper is organized as follows. In Sec. \ref{sec.model}, we set up the mixed-dimensional $t$-$J$ model. In Sec. \ref{sec.eigenstates}, we determine the low-energy single- and two-hole eigenstates. In Sec. \ref{sec.noneq_dynamics}, we study the nonequilibrium quench dynamics of two holes, before we conclude in Sec. \ref{sec.conclusions}. Throughout the paper, we set the reduced Planck constant, $\hbar$, and the lattice spacing to $1$.

\section{Model} \label{sec.model}
We consider a system of spin-$1/2$ particles placed along a two-leg ladder, described by a mixed-dimensional $t$--$J$ model with Hamiltonian $\Ham = \Ham_t + \Ham_J$ [Fig. \ref{fig.introfig}(a)]. The spins $\sigma = \uparrow,\downarrow$ can hop to nearest neighbors along the legs $\mu = 1,2$, 
\begin{align}
\Ham_t = - t\sum_{j,\sigma,\mu} \left[\tilde{c}^\dagger_{j,\mu,\sigma} \tilde{c}_{j + 1,\mu,\sigma} + \tilde{c}^\dagger_{j + 1,\mu,\sigma}\tilde{c}_{j,\mu,\sigma}\right],
\label{eq.H_t}
\end{align}
under the constraint that there is at most a single spin on each site. This is enforced by the modified particle operator $\tilde{c}^\dagger_{j,\mu,\sigma} = \hat{c}^\dagger_{j,\mu,\sigma} (1 - \hat{n}_{\mu,j})$, with $\hat{n}_{\mu,j} = \sum_\sigma \hat{c}^\dagger_{j,\mu,\sigma} \hat{c}_{j,\mu,\sigma}$ the local density operator. The nearest neighbor antiferromagnetic spin-spin coupling is assumed to be fully polarized in the $z$-direction
\begin{align}
\Ham_J = & J_{\parallel}\sum_{j,\mu} \left[\hat{S}^{(z)}_{j,\mu}\hat{S}^{(z)}_{j + 1,\mu} - \frac{\hat{n}_{j,\mu}\hat{n}_{j + 1,\mu}}{4}\right] \nn \\
+ & J_{\perp}\sum_{j} \left[\hat{S}^{(z)}_{j,1}\hat{S}^{(z)}_{j,2} - \frac{\hat{n}_{j,1}\hat{n}_{j,2}}{4}\right],
\label{eq.H_J}
\end{align}
with $J_\perp,J_\parallel > 0$. Such mixed-dimensional models \cite{Grusdt2018,Grusdt2020} have recently been proposed to yield strong binding of holes through an emergent confining string potential of overturned spins \cite{Bohrdt2022}, and was recently implemented successfully in the case of fully symmetric spin couplings \cite{Hirthe2023}. The polarized Ising-type interaction explored here, enables us to derive exact results for low-energy single- and two-hole eigenstates, as well as the full nonequilibrium dynamics of two initially localized holes. 

\section{Low-energy eigenstates} \label{sec.eigenstates}
In the absence of holes, the polarized AFM coupling in Eq. \eqref{eq.H_J} results in a perfect Ne{\'e}l ordered state, $\ket{\AF}$, for any values of $J_\parallel,J_\perp > 0$. For periodic boundary conditions of $N$ spins along each of the two legs, this results in the ground state energy
\begin{align}
E_0 = - N \frac{J_\parallel + J_\perp}{2},
\label{eq.E_0_no_holes}
\end{align}
owing to a nearest neighbor spin bond energy of $J_{\parallel} / 2$ ($J_{\perp} / 2$) along (across) the ladder. This should be contrasted to the case of isotropic spin couplings, in which case there is a quantum phase transition between Ne{\'e}l order along the ladder and spin singlet formation along the rungs as $J_\perp / J_\parallel$ is increased \cite{hida1992,sandvik1994,scalettar1994,sandvik1995,chubukov1995,Gall2021}. Utilizing this simplification, we can find the single-hole and two-hole ground states. To have a more efficient description, we employ a Holstein-Primakoff transformation and describe the system in terms of holes, $\hat{h}$, and bosonic spin excitations $\hat{s}$. The latter operators are defined with respect to the antiferromagnetic ground state $\hat{s}\ket{\AF} = 0$. The hopping Hamiltonian then reads \cite{Nielsen2023}
\begin{align}
& \Ham_t = t \sum_{j,\mu} \! \Big[ \hat{h}^\dagger_{j,\mu} F(\hat{h}_{j,\mu}, \hat{s}_{j,\mu}) F(\hat{h}_{j+1,\mu}, \hat{s}_{j+1,\mu}) \hat{h}_{j+1,\mu} \hat{s}_{j,\mu} \nn \\
& + \hat{s}^\dagger_{j + 1, \mu}\hat{h}^\dagger_{j,\mu} F(\hat{h}_{j,\mu}, \hat{s}_{j,\mu}) F(\hat{h}_{j+1,\mu}, \hat{s}_{j+1,\mu}) \hat{h}_{j+1,\mu} \Big] + {\rm H.c.}
\label{eq.H_t_holstein_primakoff}
\end{align}
Here, $F(\hat{h}, \hat{s}) = \sqrt{1 - \hat{h}^\dagger\hat{h} - \hat{s}^\dagger\hat{s}}$ ensures that there is at most a single spin excitation and a single hole on each site. The spin-coupling Hamiltonian likewise becomes
\begin{align}
\hat{H}_J = -J_{\parallel} \sum_{j,\mu} & \Big[\left(\frac{1}{2} - \hat{s}^\dagger_{j,\mu}\hat{s}_{j,\mu}\right)\left(\frac{1}{2} - \hat{s}^\dagger_{j+1,\mu}\hat{s}_{j+1,\mu}\right) + \frac{1}{4} \Big] \nn \\
\times & \big[1-\hat{h}_{j,\mu}^\dagger \hat{h}_{j,\mu}] [1-\hat{h}_{j+1,\mu}^\dagger \hat{h}_{j+1,\mu}\big] \nn \\
-J_{\perp} \sum_{j} & \Big[\left(\frac{1}{2} - \hat{s}^\dagger_{j,1}\hat{s}_{j,1}\right)\left(\frac{1}{2} - \hat{s}^\dagger_{j,2}\hat{s}_{j,2}\right) + \frac{1}{4} \Big] \nn \\
\times & \big[1-\hat{h}_{j,1}^\dagger \hat{h}_{j,1}] [1-\hat{h}_{j,2}^\dagger \hat{h}_{j,2}\big].
\label{eq.H_J_holstein_primakoff}
\end{align}
We emphasize that the spins can both be fermions and hard-core \emph{bosons}. In fact, if the holes sit on separate legs, they are distinguishable by which leg they move in. If they move along the same leg, they are equivalently distinguishable by which one is to the left and which one is to the right.  As a result, the statistics never come into play in what follows, only the hard-core constraint and the one-dimensional nature of the motion. The results, therefore, apply equally well to fermionic and hard-core bosonic spins, as one might expect from the general duality of fermions and imprenatable bosons in one dimension \cite{Girardeau1960}.

\subsection{Single-hole eigenstates}
Central to the analysis of a single hole is the insight that a single hole doped into the two-leg antiferromagnetic Ising ladder is localized. Due to inversion symmetry, the low-energy eigenstates may then be written as (assuming that the hole resides in leg 1)
\begin{align}
&\ket{\Psi_1} = \Big[C^{(0)} \hat{h}^\dagger_{0,1} + \frac{C^{(1)}}{\sqrt{2}} \big(\hat{h}^\dagger_{-1,1} + \hat{h}^\dagger_{+1,1} \big) \hat{s}^\dagger_{0} + \dots \Big]\!\ket{\AF} \nn \\
&= \Big[C^{(0)} \hat{h}^\dagger_{0,1} \!+\! \sum_{d = 1}^{N / 2} \!\frac{C^{(d)}}{\sqrt{2}} \Big(\hat{h}^\dagger_{-d,1} \!\!\!\prod_{j = 0}^{-d + 1} \!\!\!\hat{s}^\dagger_{j,1} \!+\! \hat{h}^\dagger_{d,1} \!\!\prod_{j = 0}^{d - 1}\! \hat{s}^\dagger_{j,1}\Big) \Big]\!\ket{\AF}.
\label{eq.wave_function_1hole}
\end{align}
This describes that for hole positions $d$ sites away from the central site, with total amplitude $C^{(d)}$, a resulting string of overturned spins of length $d$ appears. Taking the energy of a single \emph{stationary} hole, $E_0 + J_\parallel + J_\perp / 2$ as reference, and utilizing the Schr{\"o}dinger equation, $\Ham\ket{\Psi_1} = E_1\ket{\Psi_1}$, we obtain the equations of motion
\begin{align}
E_1 C^{(0)} &= \sqrt{2}t \cdot C^{(1)} \nn \\
E_1 C^{(1)} &= V_1(1) \cdot C^{(1)} + \sqrt{2}t \cdot C^{(1)} + t \cdot C^{(2)} \nn \\
E_1 C^{(d)} &= V_1(d) \cdot C^{(d)} + t \cdot C^{(d - 1)} + t \cdot C^{(d + 1)}.
\label{eq.EOM_1h}
\end{align}

%%%%%%%%%%%%%%%%%%%%%%%%%%%%%%%%%%%%%%%%%%%%%%%%%%%%%%%%%%%%%%%%%%% 
\begin{figure}[t!]
\begin{center}
\includegraphics[width=0.9\columnwidth]{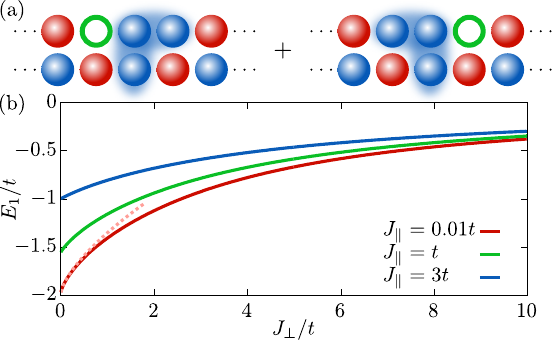}
\end{center}\vspace{-0.5cm}
\caption{(a) The hole is localized around a particular site along the ladder (top). As the hole moves, spins align in its wake, generating more and more spin frustrations [shaded blue and red background]. (b) Resulting lowest single-hole energy for indicated values of intraleg spin coupling $J_\parallel$. For $J_\perp = 0.1t$, we also show the strong coupling result [Eq. \eqref{eq.single_hole_asymptotic_energy}, red dashed line] valid for $J_\perp, J_\parallel \ll 1$. }
\label{fig.single_hole} 
\vspace{-0.25cm}
\end{figure} 
%%%%%%%%%%%%%%%%%%%%%%%%%%%%%%%%%%%%%%%%%%%%%%%%%%%%%%%%%%%%%%%%%%%

The lower equation applies for $d\geq 2$. The motion of the hole $d$ sites away leaves behind a single frustrated spin bond in leg $1$, as well as $d$ frustrated spin bonds across the ladder. This results in the linear string potential
\begin{equation}
V_1(d) = \frac{J_\parallel}{2} + d \cdot \frac{J_\perp}{2},
\label{eq.V_J_1hole}
\end{equation}
confining the hole around to its origin. The obtained equations of motion are identical in form to the recently obtained exact results in general Bethe lattices \cite{Nielsen2022_1}. Utilizing the same techniques for solving the equations of motion in Eq. \eqref{eq.EOM_1h}, without loss of generality we seek for a recursive structure of the amplitudes
\begin{align}
C^{(d + 1)} = t f^{(d + 1)}_1(E_1) C^{(d)},
\end{align}
for $d \geq 1$. Inserting this into Eq. \eqref{eq.EOM_1h}, we obtain the self-consistency equation
\begin{align}
f_1(E) = \frac{1}{E - t^2 f_1\left(E - J_\perp / 2\right)},
\label{eq.f_1_self_consistency}
\end{align}
for $d \geq 1$. Here, we have defined $f_1(E) = f_1^{(d)}(E + V_1(d))$ for a yet to be determined function $f_1$. As Eq. \eqref{eq.f_1_self_consistency} is independent of the distance $d$, $f_1(E)$ is as well. The self-consistency equation \eqref{eq.f_1_self_consistency} can, finally, be used to find a closed-form expression of $f_1(E)$ in terms of Bessel functions of the first kind, $J_\nu(x)$,
\begin{align}
f_1(E) = -\frac{1}{t}\frac{J_{\Omega(E)}\left(\frac{4t}{J_\perp}\right)}{J_{\Omega(E) - 1}\left(\frac{4t}{J_\perp}\right)}, 
\label{eq.f_1_Bessel_expression}
\end{align}
with $\Omega(E) = - 2E / J_\perp$, similar to the results in Refs. \cite{Nielsen2022_1,Chernyshev1999,Wrzosek2021}. Inserting $f_1^{(2)}(E_1) = f_1(E - V_1(2))$ in the equation for $d = 1$ in Eq. \eqref{eq.EOM_1h}, hereby, yields $C^{(1)} = \sqrt{2}t \cdot f_1(E - V_1(1)) C^{(0)}$. Inserting this into the equation for $d = 0$ in Eq. \eqref{eq.EOM_1h}, then finally results in the equation for the single-hole energy,
\begin{equation}
E_1 = \Sigma_1(E_1) = 2t^2 \cdot f_1\left(E_1 - \frac{J_\perp + J_\parallel}{2}\right),
\label{eq.E_1_equation}
\end{equation}
hereby defining the single-hole self-energy $\Sigma_1(E)$. Equation \eqref{eq.E_1_equation} actually supports a discrete series of eigenstates similar to a single hole in a Bethe lattice \cite{Nielsen2022_1}. Here, however, our main focus is on the ground state as this is important to decipher whether two holes will bind or not. 

The recursive structure of the amplitudes along with the result in Eq. \eqref{eq.f_1_Bessel_expression}, thus, allows us to construct the full many-body eigenstate with
\begin{align}
C^{(d)} &= \sqrt{2} C^{(0)} \cdot t^d \prod_{j = 1}^d f^{(j)}_1(E_1) \nn \\
&= (-1)^d\sqrt{2Z_1} \cdot \frac{J_{\Omega(E_1 - V_1(d))}\left(\frac{4t}{J_\perp}\right)}{J_{\Omega(E_1 - V_1(1)) - 1}\left(\frac{4t}{J_\perp}\right)}. 
\label{eq.single_hole_amplitudes}
\end{align}
Here, $Z_1 = [1 - \partial_E \Sigma_1(E)|_{E = E_1}]^{-1}$ is the (quasiparticle) residue of the single-hole Green's function $[E - \Sigma_1(E)]^{-1}$ at the single-hole energy $E_1$. The result $C^{(0)} = \sqrt{Z_1}$ is derived by normalizing the wave function, $1 = \braket{\Psi_1|\Psi_1}$. \\

At strong coupling, $J_\perp, J_{\parallel} \ll t$, the hole spreads out more and more, resulting in a continuum limit. This yields the asymptotic single-hole energy
\begin{equation}
E_1 \to -2t + t a^{(0)} \left(\frac{J_\perp}{2t}\right)^{2/3} + \frac{J_\parallel}{2},
\label{eq.single_hole_asymptotic_energy}
\end{equation}
with $-a^{(0)} \simeq - 1.02$ the first zero of the derivative of the Airy function, ${\rm Ai}'(x)$. In Fig. \ref{fig.single_hole}(b), we plot the single hole energy as a function $J_\perp$ for a few indicated values of $J_\parallel$. We see good agreement between Eq. \eqref{eq.single_hole_asymptotic_energy} and the full solution of Eq. \eqref{eq.E_1_equation} for $J_\parallel = 0.01t$ and $J_\perp \ll t$. 

%%%%%%%%%%%%%%%%%%%%%%%%%%%%%%%%%%%%%%%%%%%%%%%%%%%%%%%%%%%%%%%%%%% 
\begin{figure}[t!]
\begin{center}
\includegraphics[width=1.0\columnwidth]{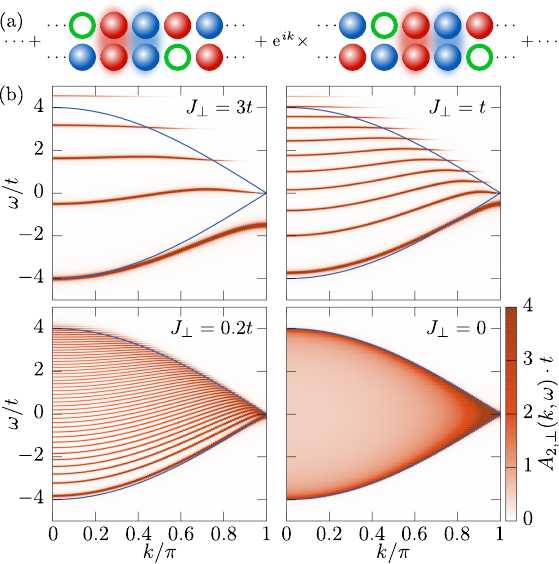}
\end{center}\vspace{-0.5cm}
\caption{(a) Two holes on separate legs are delocalized with frustrated spin bonds [shaded red and blue] between the holes. (b) Two-hole spectral function for indicated values of $J_\perp$ as a function of the crystal momentum $k$. Because of inversion symmetry, $A_{2\perp}(-k,\omega) = A_{2\perp}(+k,\omega)$, this is only plotted for $k \geq 0$. In blue is shown $\pm 4t\cos(k/2)$ for reference. The spectrum of states of the form in Eq. \eqref{eq.psi_2_k_d_separate_legs} is independent the intra-leg spin coupling $J_\parallel$.}
\label{fig.spectral_functions_separate_legs} 
\vspace{-0.25cm}
\end{figure} 
%%%%%%%%%%%%%%%%%%%%%%%%%%%%%%%%%%%%%%%%%%%%%%%%%%%%%%%%%%%%%%%%%%%

%%%%%%%%%%%%%%%%%%%%%%%%%%%%%%%%%%%%%%%%%%%%%%%%%%%%%%%%%%%%%%%%%%% 
\begin{figure}[t!]
\begin{center}
\includegraphics[width=1.0\columnwidth]{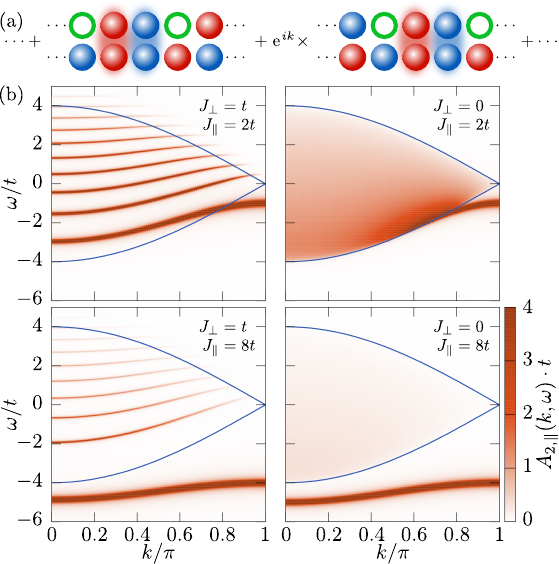}
\end{center}\vspace{-0.5cm}
\caption{(a) Like holes on separate legs, two holes on the same leg feature delocalized two-hole eigenstates and frustrated spin bonds [shaded red and blue] between the holes. (b) Two-hole spectral function for indicated values of $J_\perp$ and $J_\parallel$ as a function of the crystal momentum $k$. In blue is shown $\pm 4t\cos(k/2)$ for reference. The spectrum, here, depends on both the trans- ($J_\perp$) and intra-leg ($J_\parallel$) spin couplings. For $J_\perp \to 0$ (right), a quasiparticle band appears below the continuum, when $J_\parallel \geq 4t\cos(k/2)$. For $J_\parallel = 2t$, this corresponds to $k \geq 2\pi / 3$. }
\label{fig.spectral_functions_same_leg} 
\vspace{-0.25cm}
\end{figure} 
%%%%%%%%%%%%%%%%%%%%%%%%%%%%%%%%%%%%%%%%%%%%%%%%%%%%%%%%%%%%%%%%%%%

\subsection{Two-hole eigenstates} \label{subsec.two_hole_eigenstates}
We now focus on the low-energy two-hole eigenstates. We both consider holes moving on separate legs [Fig. \ref{fig.spectral_functions_separate_legs}], as well as holes moving along the same leg [Fig. \ref{fig.spectral_functions_same_leg}]. For holes traveling on separate legs, the breaking of spin bonds within a leg can be completely avoided by starting from a configuration of spins in which the spins to the right of the holes is moved by one lattice site to the right. Hence, if the holes move alongside each other, the perfect Ne{\'e}l order is retained and no spin bonds are broken. The appropriate two-hole eigenstates are, therefore, delocalized along the ladder. We, thus, define the states
\begin{align}
\ket{\Psi_{2\perp}(k,d)} =&\, \frac{1}{\sqrt{N}} \! \sum_j \te^{ikj} \te^{ikd/2} \hat{h}^\dagger_{j,1}\hat{h}^\dagger_{j + d,2} \nn \\
&\times\!\prod_{l > j} \hat{s}^\dagger_{l,1} \!\!\!\!\!\prod_{m > j + d} \!\!\!\!\!\hat{s}^\dagger_{m,2}\!\ket{\AF}, 
\label{eq.psi_2_k_d_separate_legs}
\end{align}
for a linear distance $d$ between the two holes. Here, we assume $N$ sites in each leg and periodic boundary conditions. In this manner, $d > 0$ ($d < 0$) indicates that the hole in leg 2 has moved $|d|$ sites to the right (left). The appearance of the string operator, $\prod_{l > j} \hat{s}^\dagger_{l,1} \prod_{m > j + d}\hat{s}^\dagger_{m,2}$, is due to the shift of the underlying AFM order by one lattice site to the right at $j$ and $j + d$.  These states have crystal momentum $k \in (-\pi, \pi]$, as translating the holes and spin strings $\hat{h}^\dagger_{j,1}\hat{h}^\dagger_{j + d,2}\prod_{l > j} \hat{s}^\dagger_{l,1} \prod_{m > j + d}\hat{s}^\dagger_{m,2}\!\ket{\AF}$ by one lattice site to the right results in an additional phase of $\te^{-ik}$. As no spin frustration within a leg occurs for this configuration of holes, the resulting low-energy eigenstates are independent of the intra-leg spin coupling $J_\parallel$.

For holes moving along the same leg, the most favorable configuration of the spins is now obtained by taking out two adjacent spins. Once again, if the holes move alongside each other, no spin bonds are broken. The delocalized states for a distance $d$ between the holes in this case, therefore, becomes
\begin{align}
\ket{\Psi_{2\parallel}(k,d)} =&\, \frac{1}{\sqrt{N}} \! \sum_j \te^{ikj} \te^{ikd/2} \hat{h}^\dagger_{j,1}\hat{h}^\dagger_{j + d,1} \!\!\!\!\prod_{l = j + 1}^{j + d - 1} \!\!\!\hat{s}^\dagger_{l,1} \!\ket{\AF}, 
\label{eq.psi_2_k_d_same_leg}
\end{align}
in which we see that a string of overturned spins forms between the two holes. Since two holes cannot sit on top of each other, let alone pass through one another, the distance is now always greater than one lattice site, $d\geq 1$. The full two-hole eigenstate can, hereby, be written as
\begin{align}
\ket{\Psi_{2s}^{(n)}(k)} = \sum_d C^{(n)}_{s}(k,d) \ket{\Psi_{2s}(k,d)},
\label{eq.psi_2_k}
\end{align}
where $s = \perp,\parallel$ denotes whether the holes move on separate legs ($\perp$) or along the same leg ($\parallel$). Additionally, the band index $n = 0,1,2,\dots$ specifies that a discrete series of two-hole bands emerge, which will become essential for describing the nonequilibrium dynamics in Sec. \ref{sec.noneq_dynamics}. The normalization condition is $1 = \braket{\Psi^{(n)}_{2s}(k)|\Psi^{(n)}_{2s}(k)} = \sum_d |C^{(n)}_{s}(k,d)|^2$. Crucially, the hopping Hamiltonian only couples the states within Eqs. \eqref{eq.psi_2_k_d_separate_legs} and \eqref{eq.psi_2_k_d_same_leg} in a well-defined hierachy. In particular, it couples holes moving on separate legs as follows: $\ket{\Psi^{(n)}_{2\perp}(k,0)} \leftrightarrow \ket{\Psi^{(n)}_{2\perp}(k,\pm 1)} \leftrightarrow \ket{\Psi^{(n)}_{2\perp}(k,\pm 2)} \cdots$. Investigating the Schr{\"o}dinger equation, $E_{2s}^{(n)}(k)\ket{\Psi_{2s}^{(n)}(k)} = \Ham \ket{\Psi_{2s}^{(n)}(k)}$ for holes on separate legs ($s = \perp$) and on on the same leg ($s = \parallel$) leads to the equations of motion
\begin{widetext}
\begin{align}
&E^{(n)}_{2\perp}(k) C^{(n)}_\perp(k,d) = V_{2\perp}(d) C^{(n)}_\perp(k,d) + 2t\cos\left(\frac{k}{2}\right)\!\!\left[C^{(n)}_\perp(k,d-1) + C^{(n)}_\perp(k,d+1)\right], 
\label{eq.EOM_2holes_separate_legs}
\end{align}
\begin{align}
&E^{(n)}_{2\parallel}(k) C^{(n)}_\parallel(k,1) = V_{2\parallel}(1) C^{(n)}_\parallel(k,1) + 2t\cos\left(\frac{k}{2}\right)C^{(n)}_\parallel(k,2), \nn \\
&E_{2\parallel}^{(n)}(k) C^{(n)}_\parallel(k,d) = V_{2\parallel}(d) C^{(n)}_\parallel(k,d) + 2t\cos\left(\frac{k}{2}\right)\left[C^{(n)}_\parallel(k,d-1) + C^{(n)}_\parallel(k,d+1)\right],
\label{eq.EOM_2holes_same_leg}
\end{align}
\end{widetext}
where the lower line in Eq. \eqref{eq.EOM_2holes_same_leg} applies for $d \geq 2$. Here, the hopping Hamiltonian couples $\ket{\Psi_{2s}^{(d)}(k)}$ and $\ket{\Psi_{2s}^{(d + 1)}(k)}$ via two pathways. For holes on separate legs, this corresponds to the hole in leg 2 hopping to the right with amplitude $t\te^{-ik/2}$, and the hole in leg 1 hopping to the left with amplitude $t\te^{+ik/2}$. For holes on the same leg, it similarly corresponds to the hole to the right to hop further to the right with amplitude $t\te^{-ik/2}$, and the hole to the left to hop further to the left with amplitude $t\te^{+ik/2}$. In any case, the associated quantum interference of these pathways leads to the total coupling of $t(\te^{-ik/2} + \te^{+ik/2}) = 2t\cos(k/2)$, as was also recognized previously \cite{Bohrdt2022}. Here, we define the two-hole linear string potentials
\begin{align}
V_{2\perp}(d) &= \left(|d| - 1\right) \cdot \frac{J_\perp}{2}, \nn \\
V_{2\parallel}(d) &= \left(d - 1\right) \cdot \frac{J_\perp}{2} - \delta_{d,1}\frac{J_\parallel}{2},
\label{eq.V_J_2holes}
\end{align}
using the energy of two separate stationary holes, $E_0 + J_\perp + 2J_\parallel$, as reference. We emphasize that for holes traveling on separate legs, the string potential does not contain the spin coupling along the ladder $J_\parallel$, because no intra-leg spin bond is broken in this case. For holes moving along the same leg, the intra-leg spin coupling $J_\parallel$ only appears at $d = 1$, as the two holes here share a frustrated intra-leg spin bond. Similar to the single-hole case, we propose the recursion relations
\begin{align}
C^{(n)}_s(k,d+1) = 2t\cos\!\left(\frac{k}{2}\right) f^{(d + 1)}_{2s}(k, E^{(n)}_{2s}(k)) C^{(n)}_s(k,d),\nn \\
C^{(n)}_\perp(k,d-1) = 2t\cos\!\left(\frac{k}{2}\right) f^{(d - 1)}_{2\perp}(k, E^{(n)}_{2\perp}(k)) C^{(n)}_\perp(k,d).
\label{eq.recursion_relations_two_holes}
\end{align}
Here, the upper line applies for both configurations of holes for $d \geq 0$ ($s = \perp$) and $d \geq 1$ ($s = \parallel$). The lower line is solely for holes on separate legs in the case of $d \leq 0$. Inserting this into Eqs. \eqref{eq.EOM_2holes_separate_legs} and \eqref{eq.EOM_2holes_same_leg} results in the self-consistency equation
\begin{equation}
f_2(k, E) = \frac{1}{E - 4t^2\cos^2(k/2)\cdot f_2(k,E - J_\perp/2)},
\label{eq.f_2_self_consistency}
\end{equation}
applying both to holes on separate legs and on the same leg. Analogous to the single hole case, we set $f_2(k,E) = f^{(d)}_{2s}(k, E + V_{2s}(d))$. As Eq. \eqref{eq.f_2_self_consistency} is independent of $d$ and the hole configuration $s = \parallel, \perp$, so is $f_2$. Note that for holes on separate legs, this also means that $f^{(d)} = f^{(-d)}$, and that $C^{(n)}(k,-d) = C^{(n)}(k,d)$ as one might expect from inversion symmetry of the system. Equation \eqref{eq.f_2_self_consistency} has the exact same structure as Eq. \eqref{eq.f_1_self_consistency} for the equivalent function $f_1$ in the single-hole case. As a result, we may simply replace $t \to  2t\cos(k/2)$ to once again obtain a closed-form expression in terms of Bessel functions of the first kind
\begin{align}
f_2(E) = -\frac{1}{2t\cos(k/2)}\frac{J_{\Omega(E)}\left(\frac{8t\cos(k/2)}{J_\perp}\right)}{J_{\Omega(E) - 1}\left(\frac{8t\cos(k/2)}{J_\perp}\right)}, 
\label{eq.f_2_Bessel_expression}
\end{align}
still with $\Omega(E) = - 2E / J_\perp$. Insertion in the equation of motion for $d = 0$ in Eq. \eqref{eq.EOM_2holes_separate_legs} and for $d = 1$ in Eq. \eqref{eq.EOM_2holes_same_leg} reveals the equations for the two-hole energies
\begin{align}
\!\!\!\! E^{(n)}_{2\perp}(k) &= -\frac{J_\perp}{2} \!+\! 8t^2 \!\cos^2\!\left(\frac{k}{2}\right) \!f_2(k, E^{(n)}_2(k)), \nn\\
\!\!\!\! E^{(n)}_{2\parallel}(k) &= -\frac{J_\parallel}{2} \!+\! 4t^2 \!\cos^2\!\left(\frac{k}{2}\right) \!f_2(k, E^{(n)}_2(k) - J_\perp / 2).\!
\label{eq.E_2_equation}
\end{align}
As for the single-hole case, we can use the recursion relations in Eq. \eqref{eq.recursion_relations_two_holes} along with Eq. \eqref{eq.f_2_Bessel_expression} to explicitly write the coefficient of the many-body wave function. For holes on separate legs, we get
\begin{align}
&C^{(n)}_\perp(k,d) = C^{(n)}_\perp(k,0)\!\left[2t\cos\!\left(\frac{k}{2}\right)\right]^{\!|d|} \!\prod_{j = 1}^{|d|} f^{(j)}_{2\perp}(k, E^{(n)}_{2\perp}(k)) \nn \\
&= (-1)^{d}\sqrt{Z^{(n)}_{2\perp}\!(k)}\frac{J_{\Omega(E^{(n)}_{2\perp}(k)) + |d| - 1}\!\left(\frac{8t\cos(k/2)}{J_\perp}\right)}{J_{\Omega(E^{(n)}_{2\perp}(k)) - 1}\!\left(\frac{8t\cos(k/2)}{J_\perp}\right)}.\!
\label{eq.two_hole_amplitudes_separate_legs}
\end{align}
for $|d| \geq 1$, using $\Omega(E - V_{2\perp}(d)) = \Omega(E) + |d| - 1$. For holes on the same leg, we similarly get
\begin{align}
\!\!\!\!&C^{(n)}_\parallel(k,d) = C^{(n)}_\parallel(k,1) \!\left[2t\cos\!\left(\frac{k}{2}\right)\right]^{\!d-1} \!\!\prod_{j = 2}^{d} f^{(j)}_{2\parallel}(k, E^{(n)}_{2\parallel}(k)) \nn \\
\!\!\!\!&= (-1)^{d-1}\sqrt{Z^{(n)}_{2\parallel}(k)}\frac{J_{\Omega(E^{(n)}_{2\parallel}(k)) + d - 1}\!\left(\frac{8t\cos(k/2)}{J_\perp}\right)}{J_{\Omega(E^{(n)}_{2\parallel}(k))}\!\left(\frac{8t\cos(k/2)}{J_\perp}\right)}. 
\label{eq.two_hole_amplitudes_same_leg}
\end{align}
for $d \geq 2$. Analogous to the single-hole case, $Z^{(n)}_{2s}(k) = [1 - \partial_\omega\Sigma_{2s}(k,E)|_{\omega = E^{(n)}_{2s}(k)}]^{-1}$ is the residue of the two-hole Green's function 
\begin{align}
G_{2\perp}(k,\omega) = \frac{1}{\omega + i\eta + J_\perp / 2 - \Sigma_{2\perp}(k,\omega + i\eta)}, \nn\\
G_{2\parallel}(k,\omega) = \frac{1}{\omega + i\eta + J_\parallel / 2 - \Sigma_{2\parallel}(k,\omega + i\eta)},
\label{eq.two_hole_greens_function}
\end{align}
for $\eta = 0^+$, $\Sigma_{2\perp}(k,\omega + i\eta) = 8t^2 \cos^2(k/2)\cdot f_2(k, \omega + i\eta)$, and $\Sigma_{2\parallel}(k,\omega + i\eta) = 4t^2 \cos^2(k/2)\cdot f_2(k, \omega + i\eta - J_\perp / 2)$. The poles of $G_{2s}$, hereby, determine the spectra for states of the forms in Eqs. \eqref{eq.psi_2_k_d_separate_legs} and \eqref{eq.psi_2_k_d_same_leg}. These are all states that have a nonzero overlap with finding holes at adjacent sites with no frustrated spin bonds. Importantly, this subfamily of states contains the two-body states with the lowest energy. The spectral functions, $A_{2s}(k,\omega) = -2\Im G_{2s}(k,\omega)$ are shown in Figs. \ref{fig.spectral_functions_separate_legs} and \ref{fig.spectral_functions_same_leg} for a few indicated values of the spin couplings. From here, the discrete bands, $E^{(n)}_{2s}(k)$, with $n = 0,1,2,\dots$ are now apparent. In the limit of $J_\perp / t \to 0^+$, a continuum of states form between $\pm 4t\cos(k/2)$. Below this continuum of states, holes traveling on the same leg [Fig. \ref{fig.spectral_functions_same_leg}(b)], feature a well-defined quasiparticle state if $J_\parallel > 4t\cos(k/2)$, in which case Eq. \eqref{eq.E_2_equation} may be solved to yield
\begin{equation}
E^{(0)}_{2\parallel}(k) = -\frac{J_\parallel}{2} - \frac{8t^2\cos^2(k/2)}{J_\parallel},
\label{eq.E_2_parallel_J_perp_0_limit}
\end{equation}
ending up at $-J_\parallel/2$ at the Brillouin zone boundary, $k=\pi$. For $J_\parallel > 4t$ [bottom right in Fig. \ref{fig.spectral_functions_same_leg}(b)], this state appears for any $k$ and a full quasiparticle band remains even for $J_\perp \to 0$. For $0 < J_\parallel < 4t$, on the other hand, a quasiparticle state appears only for crystal momenta close enough to the boundary of the Brillouin zone [top right in Fig. \ref{fig.spectral_functions_same_leg}(b)]. We note that at $k = \pi$ for general $J_\perp > 0$, there seems to be an equal spacing of the bands. In fact, inspecting Eqs. \eqref{eq.EOM_2holes_separate_legs} and \eqref{eq.EOM_2holes_same_leg} we see that the two hopping pathways destructively interfere here, giving a vanishing total hopping amplitude, $2t\cos(\pi / 2) = 0$. The states are, therefore, completely immobile and their energies are, consequently, determined by the string potentials
\begin{align}
E_{2\perp}^{(n)}(\pi) &= (n - 1)\frac{J_\perp}{2}, \nn \\
E_{2\parallel}^{(0)}(\pi) &= - \frac{J_\parallel}{2}, E_{2\parallel}^{(n)}(\pi) = n \frac{J_\perp}{2},
\end{align}
where $n\geq 0$ ($n\geq 1$) in the upper (lower) line. This gives a spacing of $J_\perp / 2$ at $k = \pi$. We note that the overall structure of the spectra in Fig. \ref{fig.spectral_functions_separate_legs} are similar to the isotropic spin coupling case in the regime of $J_\perp \gg J_\parallel$, where the underlying spin lattice resides in a disordered regime of spin-singlets on each rung \cite{Bohrdt2022}. 

%%%%%%%%%%%%%%%%%%%%%%%%%%%%%%%%%%%%%%%%%%%%%%%%%%%%%%%%%%%%%%%%%%% 
\begin{figure}[t!]
\begin{center}
\includegraphics[width=1.0\columnwidth]{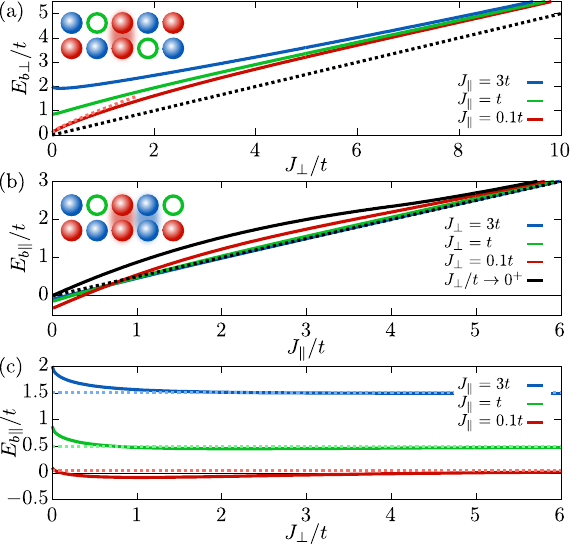}
\end{center}\vspace{-0.5cm}
\caption{(a) Trans-leg binding energy, $E_{b\perp} = 2E_1 - E_{2\perp}(0)$, versus the trans-leg spin coupling $J_\perp / t$ for several indicated values of the intra-leg spin coupling $J_\parallel$. For $J_\perp, J_\parallel \ll t$, $E_{b\perp}$ follows a $(J_\perp/t)^{2/3}$ power law behavior [light red dashed line, Eq. \eqref{eq.binding_energy_asymptotic}]. For $J_\perp \gg t$, $E_{b\perp}$ approaches $J_\perp / 2$ [black dashed line]. (b) Intra-leg binding energy, $E_{b\parallel} = 2E_1 - E_{2\parallel}(0)$, versus $J_\parallel$ for several values of $J_\perp$. $E_{b\parallel}$ approaches $J_\parallel / 2$ for $J_\parallel \gg t$ [black dashed line]. (c) Intra-leg binding energy versus $J_\perp$ instead. $E_{b\parallel}$ approaches $J_\parallel / 2$ for $J_\perp \gg t$ as well [dashed lines].  }
\label{fig.binding_energy} 
\vspace{-0.25cm}
\end{figure} 
%%%%%%%%%%%%%%%%%%%%%%%%%%%%%%%%%%%%%%%%%%%%%%%%%%%%%%%%%%%%%%%%%%%

Finally, for the lowest energy two-hole state, $k = 0$ and $n = 0$, we find that the energies at strong coupling, $J_\perp, J_\parallel \ll t$, behaves as $E^{(0)}_{2\perp}(0) = -4t(1 - a^{(0)} (J_\perp / 4t)^{2/3} / 2) + J_\perp/2$ and $E^{(0)}_{2\parallel}(0) = -4t(1 - a^{(1)} (J_\perp / 4t)^{2/3} / 2) + J_\perp/2$. Here, $-a^{(0)} \simeq -1.02$ is once again the first zero of the derivative of the Airy function [see Appendix \ref{app.continuum_limit} for details], while $-a^{(1)}\simeq -2.34$ is the first zero of the Airy function itself. Together with the single-hole energy in Eq. \eqref{eq.single_hole_asymptotic_energy}, this leads to the asymptotic binding energies, $E_{bs} = 2E_1 - E_{2s}^{(n = 0)}(k = 0)$ 
\begin{align}
E_{b\perp} \to t \cdot a^{(0)} (2 - 2^{1/3}) \left(\frac{J_\perp}{2t}\right)^{2/3} + \frac{J_\perp}{2} + J_\parallel,\nn \\
E_{b\parallel} \to t \cdot (2a^{(0)}  - 2^{1/3}a^{(1)}) \left(\frac{J_\perp}{2t}\right)^{2/3} + \frac{J_\perp}{2} + J_\parallel.
\label{eq.binding_energy_asymptotic}
\end{align}
In Fig. \ref{fig.binding_energy}(a), we plot the binding energy as a function of $J_\perp/t$ for a few indicated values of $J_\parallel$ in the case of holes on separate legs. The functional form of the binding energy in the upper line of Eq. \eqref{eq.binding_energy_asymptotic} is anticipated to remain true in the case of isotropic spin-couplings \cite{Bohrdt2022}. Together with the behavior of the binding energy for holes on the same leg [Figs. \ref{fig.binding_energy}(b)-\ref{fig.binding_energy}(c)], this lends new insights into when holes can bind strongly or not. In general, the two holes are confined by the string of overturned spins between them. This results in the dominant energy-scaling of $(J_\perp / t)^{2/3}$, and leads to a strong binding mechanism for holes on separate legs. For holes on the same leg, however, since the prefactor in front of this term is negative, $2a^{(0)} - 2^{1/3}a^{(1)} \simeq -0.90 < 0$, two holes on the same leg actually energetically prefer to unbind. Similar to recent cold-atom experiments with isotropic spin couplings \cite{Hirthe2023}, this difference can be understood from a Pauli repulsion effect. In fact, the hard-core constraint means that the boundary condition at $d = 0$ is altered from being soft for holes on separate legs to exactly zero for holes on the same leg. This results in the different prefactors of $a^{(0)} \simeq 1.02$ and $a^{(1)} \simeq 2.34$ in the two cases, which will become apparent when we investigate the spatial distribution of the holes below. We may note, however, that already for moderate values of the intra-leg spin-coupling $J_\parallel$, this unbinding is overcome and eventually reaches $J_\parallel / 2$ for $J_\parallel \gg t$. In fact, in the extreme limit of $J_\perp / t \to 0^+$ Eq. \eqref{eq.E_2_parallel_J_perp_0_limit} in combination with $E^{(0)}_{2\parallel}(0) = -4t$ for $J_\parallel / t < 4t$, results in the positive binding energy
\begin{align}
E_{b\parallel} &= J_\parallel + 4t - \sqrt{(4t)^2 + J_\parallel^2}, \, J_\parallel < 4t, \nn \\
E_{b\parallel} &= \frac{3J_\parallel}{2} + \frac{8t^2}{J_\parallel} - \sqrt{(4t)^2 + J_\parallel^2}, \, J_\parallel \geq 4t,
\end{align}
shown with a black line in Fig. \ref{fig.binding_energy}(b). Here, we use Eq. \eqref{eq.E_2_parallel_J_perp_0_limit} for $J_\parallel < 4t$ and $E_{2\parallel}(0) = -4t$ for $J_\parallel \geq 4t$, as well as $E_1 = J_\parallel/2 - 2t\sqrt{1 + (J_\parallel / 4t)^2}$ by solving Eqs. \eqref{eq.f_1_self_consistency} and \eqref{eq.E_1_equation} for $J_\perp \to 0^+$. Hence, in this limit the binding energy interpolates between two linear behaviors in the intra-leg spin coupling, from an initial $J_\parallel$ to $J_\parallel / 2$ behavior. This illustrates that two holes on the same leg bind \emph{unless} both the trans- and intra-leg spin couplings are small. Furthermore, we stress once again that these results, including the unbinding mechanism for holes on the same leg, ensues regardless of the statistics of the spins and only depends on the hard-core constraint, as one should also expect for in a system with one-dimensional motion \cite{Girardeau1960}.

In this manner, we have given a detailed account of the low-energy behavior for both intra- and trans-leg configurations. Holes on separate legs always bind with a super-linear scaling of $t \cdot (J_\perp / t)^{2/3}$ for $J_\perp, J_\parallel \ll t$. For holes on the same leg, however, the hard-core constraint results in an energy cost proportional to $t \cdot (J_\perp / t)^{2/3}$ for low $J_\perp, J_\parallel$ and leads to unbinding in this regime. However, for higher values of either spin coupling the holes will, once again, bind.
%%%%%%%%%%%%%%%%%%%%%%%%%%%%%%%%%%%%%%%%%%%%%%%%%%%%%%%%%%%%%%%%%%% 
\begin{figure}[t!]
\begin{center}
\includegraphics[width=1.0\columnwidth]{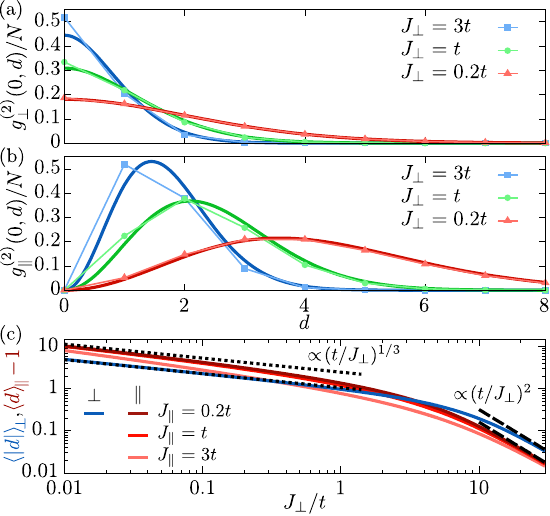}
\end{center}\vspace{-0.5cm}
\caption{(a) Trans-leg $g^{(2)}_\perp$ correlation function versus the relative distance $d$ for holes on separate legs and indicated values of the trans-leg spin coupling $J_\perp$. In dark red, green, and blue is shown the continuum limit result valid for $J_\perp / t \ll 1$ [Eq. \eqref{eq.g_2_correlator_continuum_limit}]. (b) Intra-leg $g^{(2)}_\parallel$ correlation function versus $d$ for holes on the same leg for intra-leg spin coupling $J_\parallel = 0.2t$ and indicated values of $J_\perp$. (c) Average distances $\braket{|d|}_s = \sum_d |d| g^{(2)}_s(0,d) / N$ between the two holes for the ground state at $k = 0$ as a function of $J_\perp / t$ on a log-log plot for the intra- [$s = \parallel$, red lines] and trans-leg [$s = \perp$, blue line] configurations of the holes. The intra-leg case is shown for several indicates values of $J_\parallel$. We also show the strong-correlation scaling $(t/J_\perp)^{1/3}$ [black short dashes], as well as the weak-correlation results $\propto(t/J_\perp)^{2}$ [black long dashes]. }
\label{fig.g_2_correlator} 
\vspace{-0.25cm}
\end{figure} 
%%%%%%%%%%%%%%%%%%%%%%%%%%%%%%%%%%%%%%%%%%%%%%%%%%%%%%%%%%%%%%%%%%%

Whereas a determination of the two-hole binding energy is direct proof of their ability to bind, it is simultaneously notoriously difficult to measure in modern quantum simulation experiments with ultracold atoms in synthetic lattices, such as optical lattices and Rydberg arrays. The simple reason is that the required spectroscopy entails single atom detection in e.g. time of flight or rather advanced band-mapping techniques \cite{Bohrdt2018,Brown2020a}. On the other hand, the combination of the lattice experiments and the development of quantum gas microscopy has enabled the direct and precise measurement of spatial correlations, and has successfully been employed to measure antiferromagnetic correlations in Fermi-Hubbard systems \cite{Boll2016,Mazurenko2017}, as well as characterizing the spatial properties \cite{Koepsell2019} and formation dynamics of magnetic polarons \cite{Ji2021} in such systems. For two holes, the two-point hole-hole correlators
\begin{align}
\!\!\!g^{(2)}_\perp\!(k,d) &= \frac{\braket{\hat{h}^\dagger_{1,j}\hat{h}_{1,j}\hat{h}^\dagger_{2,j+d}\hat{h}_{2,j+d}}_{\!k}}{\braket{\hat{h}^\dagger_{1,j}\hat{h}_{1,j}}_{\!k} \!\!\braket{\hat{h}^\dagger_{2,j+d}\hat{h}_{2,j+d}}_{\!k}} = N |C^{(0)}_\perp(k,d)|^2\!,\! \nn \\
\!\!\!g^{(2)}_\parallel\!(k,d) &= \frac{\braket{\hat{h}^\dagger_{1,j}\hat{h}_{1,j}\hat{h}^\dagger_{1,j+d}\hat{h}_{1,j+d}}_{\!k}}{\braket{\hat{h}^\dagger_{1,j}\hat{h}_{1,j}}_{\!k} \!\!\braket{\hat{h}^\dagger_{1,j+d}\hat{h}_{1,j+d}}_{\!k}} = N |C^{(0)}_\parallel(k,d)|^2\!,\!
\label{eq.g_2_correlator}
\end{align}
provide such a spatial probe of their binding, as was also recently used in experiments \cite{Hirthe2023}. In Eq. \eqref{eq.g_2_correlator}, the average is taken for the states $\ket{\Psi_{2s}^{(0)}(k)}$ with $s = \perp,\parallel$ in Eq. \eqref{eq.psi_2_k} residing in the lowest band $E_{2s}^{(0)}(k)$. We utilize that the amplitude $C^{(0)}_s(k,d)$ gives the probability to observe the holes at distance $d$, $|C^{(0)}_s(k,d)|^2$. Therefore, the numerator simply gives $|C^{(0)}_s(k,d)|^2 / N$, whereas the uniform spreading of the holes means that $\braket{\hat{h}^\dagger_{\mu,j}\hat{h}_{\mu,j}}_{k} = 1 / N$, for both legs $\mu = 1,2$. In Figs. \ref{fig.g_2_correlator}(a)--\ref{fig.g_2_correlator}(b), we plot these correlators as a function of $d$ for several values of $J_\perp$. For lower values of $J_\perp / t$, the holes separate more and more from each other as one expects for a higher mobility. We note that already for $J_\perp = 3t$, the probability of finding the holes as nearest neighbors has dropped to around $50\%$. For $J_\perp / t \ll 1$ -- and $J_\parallel / t \ll 1$ in the intra-leg case -- the relative wave functions of the holes, $C^{(n)}_s(k,d)$, can be mapped to a continuum model. In this limit, they fulfill a continuous one-dimensional Schr{\"o}dinger equation with a mass scaling as $t$ and a linear potential scaling with $J_\perp$ \cite{Kane1989,Zhong1995,Grusdt2018_2,Nielsen2022_2}. As a result, the relative wave functions takes on the form of Airy functions [see Appendix \ref{app.continuum_limit}], resulting in
\begin{align}
\frac{g^{(2)}_\perp(k,d)}{N} \to A_0^2 \lambda(k) [{\rm Ai}(\lambda(k)|d| - a^{(0)})]^2, \nn \\
\frac{g^{(2)}_\parallel(k,d)}{N} \to A_1^2 \lambda(k) [{\rm Ai}(\lambda(k)d - a^{(1)})]^2,
\label{eq.g_2_correlator_continuum_limit}
\end{align}
with the effective inverse length scale $\lambda(k) = [J_\perp/(4t\cos(k/2))]^{1/3}$, and the normalization constants $A_j$. We compare to these continuum results and see remarkably good agreement away from $d = 0$ even for relatively large values of $J_\perp$. Additionally, we show the average distance between the holes $\braket{|d|}_{sk} = \sum_d |d| g^{(2)}_s(k,d) / N$ as a function of $J_\perp / t$ in Fig. \ref{fig.g_2_correlator}(c) for the ground state at $k = 0$. This reveals the strong-correlation scaling
\begin{equation}
\braket{d}_{\parallel k} = a^{(1)}\cdot\braket{|d|}_{\perp k} = \frac{2a^{(1)}}{3\lambda(k)} \propto \left(\frac{t}{J_\perp}\right)^{1/3},
\label{eq.average_distance}
\end{equation}
for $J_\perp / t \ll 1$. Figure \ref{fig.g_2_correlator}(c) shows that this asymptotic form is already accurate for $J_\perp / t \leq 1$. We attribute this to the fact that the effective interaction strength for two holes is $4t/J_\perp$, rather than just $t/J_\perp$. For weak correlations, we similarly get $\braket{d}_{\parallel k} - 1 = \braket{|d|}_{\perp k} / 2 = 1 / \lambda^6(k)\propto (t/J_\perp)^{2}$, becoming accurate for $J_\perp / t \geq 10$ in Fig. \ref{fig.g_2_correlator}(c). Importantly, we emphasize that for holes on the same leg, the hard-core constraint $g^{(2)}_\parallel(k,d = 0) = 0$ results in a different boundary condition in the continuum limit. This change in the boundary condition alters the relative wave function from being on the form of ${\rm Ai}(\lambda(k)|d| - a^{(0)})$ to ${\rm Ai}(\lambda(k)d - a^{(1)})$, and changes the prefactor of the $(J_\perp / t)^{2/3}$ term in the two-hole energy from $a^{(0)} \simeq 1.02$ to $a^{(1)} \simeq 2.34$. This also results in a significant qualitative change in the relative spatial distribution of the holes. For holes on separate legs, the holes are always most likely to be found as nearest neigbors, whereas this is not true at all for holes on the same leg. This leads to more distant holes in the intra-leg configuration and explains the extra factor of $a^{(1)} \simeq 2.34$ in $\braket{d}_{\parallel k}$. In Fig. \ref{fig.g_2_correlator}, we focus on the ground state behavior, i.e. $k = 0$. We note, however, that as the momentum approaches the edge of the Brillouin zone, the correlator compresses more and more and eventually the holes only sit next to each other: $g^{(2)}_\perp(k = \pi,0) = N$ and $g^{(2)}_\parallel(k = \pi,1) = N$. 

%%%%%%%%%%%%%%%%%%%%%%%%%%%%%%%%%%%%%%%%%%%%%%%%%%%%%%%%%%%%%%%%%%% 
\begin{figure}[t!]
\begin{center}
\includegraphics[width=1.0\columnwidth]{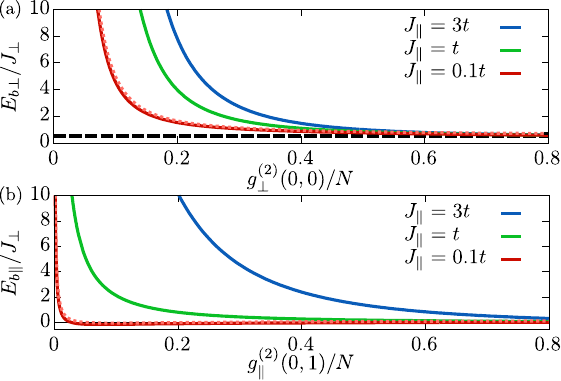}
\end{center}\vspace{-0.5cm}
\caption{(a) Trans-leg binding energy in units of $J_\perp$ versus $g^{(2)}_\perp$ at $k = 0$ and for adjacent holes, $d = 0$, for indicated values of the intraleg spin coupling $J_\parallel$. This is compared to the asymptotic behavior in Eq. \eqref{eq.Eb_vs_g2_perp} for $J_\parallel = 0.1t$ [light red dashed line], as well as the weak coupling binding energy [black long dashed line]. We observe a monotically decreasing behavior of $E_b/J_\perp$ for increasing $g^{(2)}_\perp$. (b) Intra-leg binding energy versus $g^{(2)}_\parallel$ at $k = 0$ and adjacent holes, $d = 1$, for the same values of $J_\parallel$, and also compared to the asymptotic behavior [Eq. \eqref{eq.Eb_vs_g2_parallel}]. For small $J_\parallel$, the binding energy false off very quickly with increasing $g^{(2)}_\parallel$.}
\label{fig.Eb_vs_g2} 
\vspace{-0.25cm}
\end{figure} 
%%%%%%%%%%%%%%%%%%%%%%%%%%%%%%%%%%%%%%%%%%%%%%%%%%%%%%%%%%%%%%%%%%%

Equation \eqref{eq.g_2_correlator_continuum_limit} reveals that for holes on separate legs, the correlator at $d = 0$ scales with the inverse length scale $\lambda(k)\propto (J_\perp / t)^{1/3}$. Since the binding energy scales with $t(J_\perp / t)^{2/3}$, we get that $E_{b\perp} / J_\perp \propto 1/g^{(2)}(0,0)$ at strong coupling. More precisely,
\begin{equation}
\frac{E_{b\perp}}{J_\perp} = \frac{c_\perp}{g^{(2)}_\perp(0,0) / N} + \frac{1}{2} + \frac{J_\parallel}{J_\perp},
\label{eq.Eb_vs_g2_perp}
\end{equation}
with $c_\perp = 2^{-4/3}(1 - 2^{-2/3})$ for $J_\perp, J_\parallel \ll t$. This is very valuable for quantum simulation experiments, as it provides an indirect probe of the binding energy. In fact, in Ref. \cite{Hirthe2023} an approximate relation at finite temperatures between the binding energy and the two-point correlator was used in this regard. For the configuration with two holes on the same leg, Eq. \eqref{eq.g_2_correlator_continuum_limit} similarly gives $g^{(2)}_\parallel(k,1) \propto \lambda^3(k)$. The asymptotic relation between the binding energy and the correlation function, therefore, now takes on the form
\begin{equation}
\frac{E_{b\parallel}}{J_\perp} = \frac{c_\parallel}{[g^{(2)}_\parallel(0,0) / N]^{1/3}} + \frac{1}{2} + \frac{J_\parallel}{J_\perp},
\label{eq.Eb_vs_g2_parallel}
\end{equation}
with $c_\parallel = 2^{-1/3}(a^{(0)} - 2^{-2/3}a^{(1)})$. This asymptotic relationship indicates that $g^{(2)}_\parallel(0,0)$ must be much smaller to observe an impact on the binding energy. To explore these behaviors further, we plot the binding energy versus $g^{(2)}$ in Fig. \ref{fig.Eb_vs_g2}. For holes on separate legs, this reveals a monotonic relation between the binding energy and the $g^{(2)}$ correlator for nearest neighbor holes for any value of $J_\parallel$, which indeed enables experiments to infer a binding energy from a measured $g^{(2)}$ function. In the case of holes on the same leg, however, the applicability of this approach may, however, depend quite crucially on the value of $J_\parallel$. In fact, for $J_\parallel\ll t$, we see that only at extremely low value of $g^{(2)}_\parallel(0,0)$ does the binding energy start to change significantly, which will naturally make it much harder to infer a binding energy from a measured $g^{(2)}$ function. 

%%%%%%%%%%%%%%%%%%%%%%%%%%%%%%%%%%%%%%%%%%%%%%%%%%%%%%%%%%%%%%%%%%% 
\begin{figure}[t!]
\begin{center}
\includegraphics[width=1.0\columnwidth]{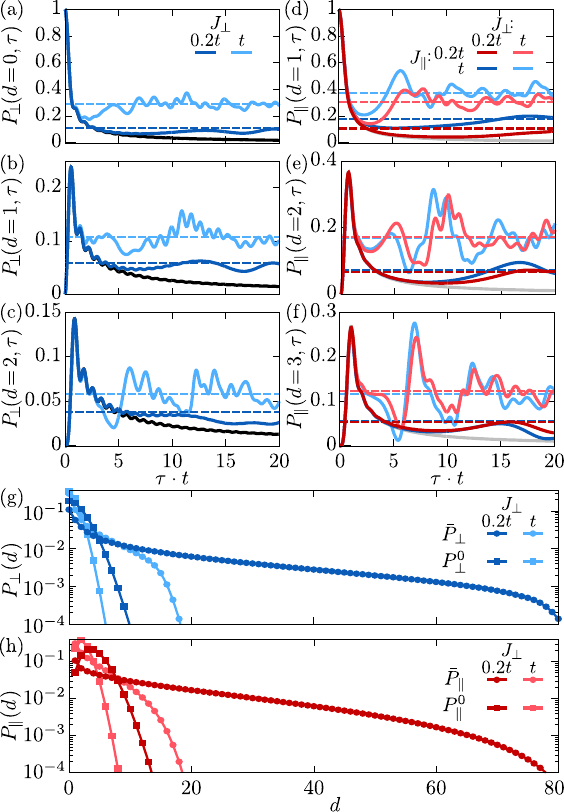}
\end{center}\vspace{-0.5cm}
\caption{Temporal evolution of the probability to find the two holes at distances $d = 0$ (a), $d = 1$ (b), $d = 2$ (c) for holes on separate legs of the ladder, and of finding the holes at distances $d = 1$ (d), $d = 2$ (e), and $d = 3$ (f) for holes on the same leg. This is shown for indicated values of the trans- ($J_\perp$) and intra-leg ($J_\parallel$) spin couplings and compared to the quantum walk for holes on separate legs [black lines] and on the same leg [grey lines]. We also show the long-time average probability distributions $\bar{P}_\perp(d), \bar{P}_\parallel(d)$ in dashed lines. (g)-(h) $\bar{P}_\perp(d), \bar{P}_\parallel(d)$ compared to the ground state probability distributions, $P^0_\perp (d) = |C^{(0)}_\perp(k = 0, d)|^2$ and $P^0_\parallel (d) = |C^{(0)}_\parallel(k = 0, d)|^2$, for holes on separate legs (g) and on the same leg (h).} 
\label{fig.prob_vs_time} 
\vspace{-0.25cm}
\end{figure} 
%%%%%%%%%%%%%%%%%%%%%%%%%%%%%%%%%%%%%%%%%%%%%%%%%%%%%%%%%%%%%%%%%%%
\section{Nonequilibrium dynamics}  \label{sec.noneq_dynamics}
In this section, we investigate the nonequilibrium dynamics of two initially localized holes. Such a quench experiment is a natural choice for quantum simulation experiments, and have recently been considered for the motion of a hole in a Fermi-Hubbard background \cite{Ji2021}, in which they were able to see the crossover dynamics from an initial free ballistic motion of the hole, signatures of string oscillations, and finally to the ballistic motion of magnetic polaron quasiparticles at long times \cite{Nielsen2022_2,Bohrdt2020}. In the current setup, we investigate the situation where the holes are localized and start out as nearest neighbor, described by the wave functions for the separate-legs ($\perp$)and same-leg ($\parallel$) configurations
\begin{align}
\!\!\!\!\ket{\Psi_{2\perp}(\tau = 0)} &= \hat{h}^\dagger_{0,1}\hat{h}^\dagger_{0,2} \!\prod_{l > 0} \hat{s}^\dagger_{l,1} \!\!\prod_{m > 0} \!\!\hat{s}^\dagger_{m,2}\!\ket{\AF} \nn \\
&= \frac{1}{\sqrt{N}}\sum_k \ket{\Psi_{2\perp}(k,0)}, \nn \\
\!\!\!\!\ket{\Psi_{2\parallel}(\tau = 0)} &= \hat{h}^\dagger_{0,1}\hat{h}^\dagger_{1,1}\!\ket{\AF} = \frac{1}{\sqrt{N}}\!\sum_k \ket{\Psi_{2\parallel}(k,1)},\!
\label{eq.initial_state_2hole_dynamics}
\end{align}
using $\tau$ as the variable for time to distinguish it from the hopping amplitude $t$. In the second line, as well as the last expression of the third line, we utilize that the initial state is the superposition of all the crystal momentum states in Eqs. \eqref{eq.psi_2_k_d_separate_legs} and \eqref{eq.psi_2_k_d_same_leg} for $d = 0$ and $d = 1$, respectively. To determine the full dynamics, we calculate the overlap of the initial states with the two-hole eigenstates in Eq. \eqref{eq.psi_2_k}
\begin{align}
\braket{\Psi^{(n)}_{2\perp}(k)|\Psi_{2\perp}(\tau = 0)} &= \frac{C_\perp^{(n)}(k,0)}{\sqrt{N}}, \nn \\
\braket{\Psi^{(n)}_{2\parallel}(k)|\Psi_{2\parallel}(\tau = 0)} &= \frac{C_\perp^{(n)}(k,1)}{\sqrt{N}} \te^{-ik/2}.
\label{eq.overlap_initial_state}
\end{align}
Since the eigenstates are delocalized over the entire lattice, there is an overall factor of $1/\sqrt{N}$, whereas the factors of $C_\perp^{(n)}(k,0) = \sqrt{Z_{2\perp}^{(n)}(k)}$ and $C_\perp^{(n)}(k,1) = \sqrt{Z_{2\parallel}^{(n)}(k)}$ are the amplitudes for finding the holes as nearest neigbhors in the states $\ket{\Psi^{(n)}_{2\perp}(k)}$ and $\ket{\Psi^{(n)}_{2\parallel}(k)}$, respectively. See also Eqs. \eqref{eq.two_hole_amplitudes_separate_legs} and \eqref{eq.two_hole_amplitudes_same_leg}. We note that it is crucial to take into account that states in all the bands $n$ have an overlap with the initial state and must be taking into account. The nonequilibrium wave functions are then
\begin{align}
\!\!\!\!\!\!\ket{\Psi_{2\perp}(\tau)} &= \sum_{k,n} \te^{-iH\tau}\ket{\Psi^{(n)}_{2\perp}(k)}\braket{\Psi^{(n)}_{2\perp}(k)|\Psi_{2\perp}(\tau = 0)} \nn \\
&= \!\!\frac{1}{\sqrt{N}}\!\sum_{k,n} \!C_\perp^{(n)}(k,0) \te^{-iE^{(n)}_{2\perp}(k)\tau}\ket{\Psi^{(n)}_{2\perp}(k)}\!,\! \label{eq.noneq_wavefunc_separate_legs}
\end{align}
for the separate-legs configuration, and 
\begin{align}
\!\!\!\!\!\!\ket{\Psi_{2\parallel}(\tau)} &= \sum_{k,n} \te^{-iH\tau}\ket{\Psi^{(n)}_{2\parallel}(k)}\braket{\Psi^{(n)}_{2\parallel}(k)|\Psi_{2\perp}(\tau = 0)} \nn \\
&= \!\!\frac{1}{\sqrt{N}}\!\sum_{k,n} \te^{-ik}C_\parallel^{(n)}\!(k,1) \te^{-iE^{(n)}_{2\parallel}(k)\tau}\!\ket{\Psi^{(n)}_{2\parallel}(k)}\!. \!\!\label{eq.noneq_wavefunc_same_leg}
\end{align}
for holes on the same leg. To describe the two-hole dynamics more concisely, we use Eqs. \eqref{eq.noneq_wavefunc_separate_legs} and \eqref{eq.noneq_wavefunc_same_leg} to compute the probability of finding the holes at a distance $d$ as a function of time
\begin{align}
\!\!\!\!\!P_\perp(d,\tau) \!&=\! \frac{1}{N} \!\sum_{k}\Big|\!\sum_n C^{(n)}_\perp(k,0) C^{(n)}_\perp(k,d) \te^{-iE^{(n)}_{2\perp}(k)\tau}\Big|^2\!\!,\!\! \nn \\
\!\!\!\!\!P_\parallel(d,\tau) \!&=\! \frac{1}{N} \!\sum_{k}\Big|\!\sum_n C^{(n)}_\parallel(k,1) C^{(n)}_\parallel(k,d) \te^{-iE^{(n)}_{2\parallel}(k)\tau}\!\Big|^2\!\!,\!\!
\label{eq.probability_distribution_distance}
\end{align}
describing the relative wave function versus time. Figures \ref{fig.prob_vs_time}(a)-\ref{fig.prob_vs_time}(f) shows the dynamics of these probability distributions for several indicated distances, $d$. At short times, the holes initially blow apart \emph{ballistically} as described by the quantum walks 
\begin{align}
\!\!\!\!\!P_\perp^{(\rm q.w.)}(d,\tau) \!&=\! \frac{1}{N} \!\sum_{k}\cos(kd)\Big|\frac{1}{N}\!\sum_p \te^{+i(\varepsilon_{p} - \varepsilon_{p + k})\tau}\Big|^2\!\!,\!\! \nn \\
\!\!\!\!\!P_\parallel^{(\rm q.w.)}(d,\tau) \!&=\! \frac{2}{N} \!\sum_{k}\cos(k)\Bigg[\cos(kd)\Big|\frac{1}{N}\!\sum_p \te^{+i(\varepsilon_{p} - \varepsilon_{p + k})\tau}\Big|^2 \nn\\
&\phantom{= \frac{2}{N} \!\sum_{k}} - \Big|\frac{1}{N}\!\sum_p \te^{ip} \te^{+i(\varepsilon_{p} - \varepsilon_{p + k})\tau}\Big|^2\Bigg],
\label{eq.probability_distribution_distance_free}
\end{align}
derived in Appendix \ref{app.quantum_walks}. For holes on the same leg, lower line in Eq. \eqref{eq.probability_distribution_distance_free}, the hard-core property of the holes constrains their motion and slightly alters it from the quantum walk of independent holes on separate legs. On longer timescales, the distribution of the holes is seen to oscillate around the time-averaged distributions
\begin{align}
\bar{P}_\perp(d) &= \lim_{T\to\infty} \frac{1}{T}\int_0^T d\tau \; P_\perp(d,\tau) \nn \\
&= \frac{1}{N}\sum_{k,n} |C_\perp^{(n)}(k,0)|^2 |C_\perp^{(n)}(k,d)|^2,\nn\\
\bar{P}_\parallel(d) &= \lim_{T\to\infty} \frac{1}{T}\int_0^T d\tau \; P_\parallel(d,\tau) \nn \\
&= \frac{1}{N}\sum_{k,n} |C_\parallel^{(n)}(k,1)|^2 |C_\parallel^{(n)}(k,d)|^2,
\label{eq.time_averaged_prob_dist}
\end{align}
which denotes the steady state approximately reached on long timescales. We note, however, that because the two-hole spectra in Figs. \ref{fig.spectral_functions_separate_legs} and \ref{fig.spectral_functions_same_leg} consists of a discrete set of coherent peaks for any nonzero value of the trans-leg spin coupling $J_\perp$, the motion will generally be highly aperiodic and never settle at its long-time average.  As a result, the system does not fully equilibrate. It does still, however, give the characterize distribution of the holes at long times. To understand this further, in Figs. \ref{fig.prob_vs_time}(g)-\ref{fig.prob_vs_time}(h), we compare it to the probability distribution for the two holes in the ground state. We observe that the behavior of the steady state is markedly different from the ground state. First and foremost, the state will dynamically extend over much larger length scales than its ground state counterpart. This is challenging for a cold-atom simulation experiment, and may hinder the observation of the long-time dynamics. However, we stress that already over a few hopping timescales $1 / t$, does the dynamics start to deviate from the quantum walk. 

%%%%%%%%%%%%%%%%%%%%%%%%%%%%%%%%%%%%%%%%%%%%%%%%%%%%%%%%%%%%%%%%%%% 
\begin{figure}[t!]
\begin{center}
\includegraphics[width=1.0\columnwidth]{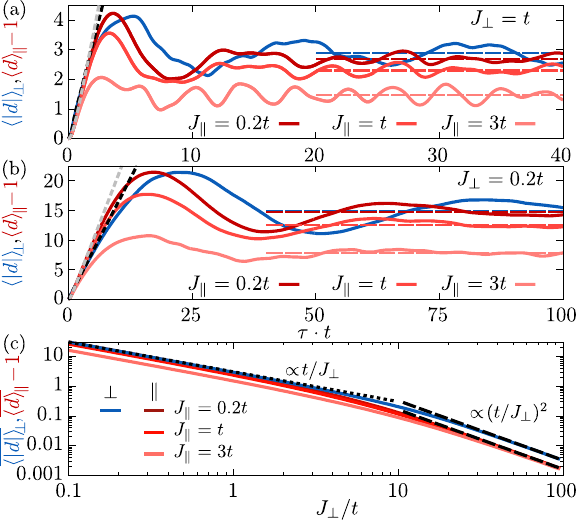}
\end{center}\vspace{-0.5cm}
\caption{(a)--(b) Mean distance versus time for indicated intra- and trans-leg spin couplings for holes on separate legs [blue lines] and the same leg [red lines]. The black and grey dashed lines show the quantum walk for holes on separate legs and the same leg, respectively. At long times, the holes oscillate around well-defined mean distances $\overline{\braket{|d|}}_\perp, \overline{\braket{d}}_\parallel$ [long-dashed lines], shown in (c) as a function of the trans-leg spin coupling $J_\perp$, for indicated values of the intra-leg spin coupling $J_\parallel$. For weak correlations, the time-averaged mean distances scale as $(t/J_\perp)^{2}$ as the eigenstates in Fig. \ref{fig.g_2_correlator}, while the scaling in the strong correlation limit is changed from $(t/J_\perp)^{1/3}$ for the eigenstates to $t/J_\perp$ for the dynamics. }
\label{fig.average_distance} 
\vspace{-0.25cm}
\end{figure} 
%%%%%%%%%%%%%%%%%%%%%%%%%%%%%%%%%%%%%%%%%%%%%%%%%%%%%%%%%%%%%%%%%%%

To investigate this more quantitatively using a simple experimental probe, we compute the average distances
\begin{align}
\braket{|d|}_\perp(\tau) &= \sum_d |d| P_\perp(d,\tau), \nn \\
\braket{d}_\parallel(\tau) &= \sum_d d P_\parallel(d,\tau), 
\end{align}
as a function of time. Two exemplary results are shown in Figs. \ref{fig.average_distance}(a)-\ref{fig.average_distance}(b). For times $\tau < 2 / t$, holes on the same leg will depart slightly slower than holes on separate legs, simply because there are more configurations available for holes on separate legs in the very first hop. After that, the hard-core constraint leads to faster divergent motion for holes on the same leg, but the motion remains a ballistic quantum walk. When the holes cross their long-time average, $\overline{\braket{|d|}}_\perp, \overline{\braket{d}}_\parallel$, the motion starts to deviate significantly from the initial ballistic behavior and instead crosses over to oscillations around $\overline{\braket{|d|}}_\perp, \overline{\braket{d}}_\parallel$. We use this to define the dynamical regimes in Fig. \ref{fig.introfig}(d). In fact, the interhole distance in the separate-legs configuration quickly evolves linearly in time, $\braket{|d|}_\perp^{(\rm q.w.)} = 13/8 (t\cdot\tau)$. We, therefore, simply define the crossover timescale in Fig. \ref{fig.introfig}(d) as the time $\tau$ at which $\braket{|d|}_\perp^{(\rm q.w.)} \simeq 13/8 (t\cdot\tau) = \overline{\braket{|d|}}_\perp$. We, hereby, note that the crossover from the quantum walk to the string oscillation regime for say $J_\perp = 3t$, happens already around $\tau \simeq 1/t$. This should be a signifant help to see at least the onset of the oscillation regime in a cold-atom simulation \cite{Ji2021}.

Figure \ref{fig.average_distance}(c), finally, shows the long-time average distances 
\begin{align}
\overline{\braket{|d|}}_\perp &= \sum_d |d| \cdot\bar{P}_\perp(d) = \frac{1}{N}\sum_{k,n} Z_{2\perp}^{(n)}(k) \braket{|d|}^{(n)}_{\perp k}, \nn \\
\overline{\braket{d}}_\parallel &= \sum_d d \cdot\bar{P}_\parallel(d) = \frac{1}{N}\sum_{k,n} Z_{2\parallel}^{(n)}(k) \braket{|d|}^{(n)}_{\parallel k}, 
\label{eq.ave_distance_long_time}
\end{align}
as a function of the trans-leg spin coupling, $J_\perp$. For the same-leg configuration, this is, furthermore, shown for indicated values of the intra-leg spin coupling, $J_\parallel$. In Equation \eqref{eq.ave_distance_long_time}, we use that the probability to find the holes as nearest neighbors in a given eigenstate $n,k$ is given by the quasiparticle residues $|C_\perp^{(n)}(k,0)|^2 = Z_{2\perp}^{(n)}(k)$, $|C_\perp^{(n)}(k,1)|^2 = Z_{2\parallel}^{(n)}(k)$. The expressions to the right in Eq. \eqref{eq.ave_distance_long_time}, hereby, reveal that the long-time averages of the nonequilibrium average distances are given by an appropriate mean value of the inter-hole average distances, $\braket{|d|}^{(n)}_{\perp k}$, $\braket{|d|}^{(n)}_{\parallel k}$, of the eigenstates. One could, therefore, na{\"i}vely expect these to scale in the same manner as the eigenstates with $t/J_\perp$. For weak correlations, $J_\perp \gg t$, this is indeed the case, where we find that this distance is the same as for the ground states in Fig. \ref{fig.g_2_correlator}(c) and, thus, vanishes as $(t/J)^2$. For strong correlations $J_\perp \ll t$, however, we see that the distance between the holes reaches a universal $t/J_\perp$-scaling. For the same-leg configuration, this also reguires $J_\parallel \ll t$. This same scaling was found for the motion of a single hole in antiferromagnetic Bethe lattice structures \cite{Nielsen2022_1}, and shows a remarkable qualitative difference to the equilibrium situation with eigenstates supporting only a much weaker $(t/J)^{1/3}$ scaling for the eigenstates, Fig. \ref{fig.g_2_correlator}(c). This quantifies the qualitative picture drawn from Figs. \ref{fig.prob_vs_time}(g)-\ref{fig.prob_vs_time}(h) that the quenched holes already for intermediate values of $J_\perp \sim t$ spread out much more than one would expect from the spatial size of the ground state. 

For the computation of the dynamics, we increase the system size $N$ and the number of included bands $n_{\rm max}$ until the results have converged. As a rule of thumb, this is achieved when the system size is a few times larger than the mean distance between the holes. For the most strongly correlated case of $J_\perp = 0.1t$, we go up to $N = 600$ sites and $n_{\rm max} = 88$ bands. Utilizing the inversion symmetry of the system, we, hereby, need to resolve the energy and residue of $N / 2 \cdot n_{\rm max} = 26400$ states. This emphasizes that we need a very thorough understanding of the eigenspectrum to be able to simulate the quench dynamics in this manner.

\section{Conclusions and outlook} \label{sec.conclusions}
Inspired by the recent experimental realization of hole pairing in a cold-atom quantum simulator \cite{Hirthe2023} of a mixed-dimensional $t$--$J$ model \cite{Bohrdt2022}, we have investigated a simplified setup of Ising spin interactions. This allowed us to determine the \emph{exact} low-energy single- and two-hole eigenstates. We used this to rigorously show that two holes on separate legs bind strongly to each other in the strongly correlated regime of $J_\perp,J_\parallel \ll t$, in that it features a \emph{superlinear} binding energy: $E_b \propto (J_\perp/t)^{2/3}$.

Furthermore, we used this exact description to rigorously account for the nonequilibrium quench dynamics following two initially localized holes at adjacent sites. Similar dynamics has previously been investigated for a single hole in a square lattice geometry \cite{Ji2021}, whose analysis provided evidence of emergent dynamical regimes, describing the crossover from a quantum walk on short timescales to string oscillations at intermediate timescales and finally the ballistic motion of magnetic polaron quasiparticles at long times \cite{Nielsen2022_2, Bohrdt2020}. In the present mixed-dimensional setup, we found a similar dichotomy of the dynamics for \emph{two} holes with two major differences. First, the holes are confined to each other, such that their distance remains finite. Second, the string oscillations in the present scenario have an infinite lifetime, and, therefore, persist indefinitely, hindering the long-time equilibration of the system.

These results pave the way for a precise comparison with state-of-the-art cold-atom quantum simulation experiments. There are three essential ingredients that makes the system interesting from this perspective. First, our mixed-dimensional model may be implemented both with fermions and hard-core bosons. Second, the effective interaction strength of $4t/J_\perp$ means that the experiments can more easily access a strongly correlated regime already for $J_\perp \lesssim 4t$. Third, this is particularly relevant for the quench dynamics, where the crossover from the quantum walk to the string oscillations already happens around times of $\tau \simeq 1 / t$ in this intermediate parameter regime. We, therefore, believe that it should be possible to experimentally access the crossover from the quantum walk to the confinement-induced oscillations.

Furthermore, such experiments also naturally lead to two interesting roads ahead. First, by systematically increasing the number of legs in the ladder, one can carefully analyze if the system supports the formation of stripes \cite{Zaanen2001,Corboz2011,Corboz2014,Hoyos2015} inherent to the phenomenology of high-temperature superconductors. Such inquiries were investigated in Ref. \cite{Grusdt2020} using quantum Monte Carlo calculations, in the special case where the trans- and intra-leg spin interactions are equal. We speculate that our methodology may lend exact insights into this scenario at zero temperature. Second, we believe that it is possible to use the present methodology also at nonzero temperatures. This would require a thorough analysis of the eigenstates as more and more spins are flipped. This would enable the exact determination of the nonequilibrium dynamics of holes at finite temperatures, and could be used to answer whether the holes will deconfine \cite{Hahn2022} from each other as a result of thermal spin fluctuations.

\acknowledgements
The author thanks Umesh Kumar, Marton Kanasz-Nagy and J. Ignacio Cirac for valuable discussions. This article was supported by the Carlsberg Foundation through a Carlsberg Internationalisation Fellowship. 

\appendix

\section{Continuum limit for two holes} \label{app.continuum_limit}
In this appendix, we derive the two-hole energy in the limit $J_\perp / t \ll 1$. The derivation is very similar to the recent results in Bethe lattice structures \cite{Nielsen2022_1}. \\

We initially analyze the situation for holes on separate legs, starting from the equations of motion in Eq. \eqref{eq.EOM_2holes_separate_legs}. Using $\psi^{(n)}(k,d) = (-1)^d C^{(n)}(k,d)$, we obtain
\begin{align}
&\frac{E^{(n)}_{2\perp}(k) + 4t\cos(k/2) + J_\perp / 2}{2t\cos(k/2)} \psi^{(n)}(k,d) = \nn \\
&\frac{J_\perp}{4t\cos(k/2)}|d| \psi^{(n)}(k,d) \nn\\
&- \left[\psi^{(n)}(k,d-1) - 2\psi^{(n)}(k,d) + \psi^{(n)}(k,d+1)\right],
\label{eq.EOM_2holes_2}
\end{align}
valid for any $k\neq \pm \pi$. We then rescale lengths according to $d = x / \lambda$, and define $\phi^{(n)}(k,x) = \psi^{(n)}(k,x / \lambda) / \sqrt{\lambda}$. Inserting this in Eq. \eqref{eq.EOM_2holes_2}, we get
\begin{align}
&a^{(n)} \phi^{(n)}(k,x) = \frac{J_\perp}{4t\cos(k/2)}\frac{|x|}{\lambda^3} \phi^{(n)}(k,x) \nn\\
&- \frac{\phi^{(n)}(k,x-\lambda) - 2\phi^{(n)}(k,x) + \phi^{(n)}(k,x+\lambda)}{\lambda^2},
\label{eq.EOM_2holes_3}
\end{align}
with $a^{(n)} = (E^{(n)}_{2\perp}(k) + 4t\cos(k/2) + J_\perp / 2) / (2t\cos(k/2)\cdot\lambda^2)$. To remove the dependency on $J_\perp/t$, we set $\lambda^3(k) = J_\perp/(4t\cos(k/2))$. In the limit of $\lambda\propto (J_\perp / t)^{1/3} \to 0^+$, the second line of Eq. \eqref{eq.EOM_2holes_3} simply becomes the second derivative of $\phi$. Hence, we are left with the differential equation
\begin{align}
&a^{(n)} \phi^{(n)}(k,x) =|x| \phi^{(n)}(k,x) - \frac{d^2\phi^{(n)}(k,x)}{dx^2},
\label{eq.EOM_2holes_4}
\end{align}
where the wave function is subject to the normalization condition $\int_{-\infty}^{+\infty} dx |\phi(k,x)|^2 = 1$. Hence, we effectively have a single particle in one dimension subject to a linear potential in this limit. Rearranging yields 
\begin{align}
\frac{d^2f(y)}{dy^2} = y f(y),
\label{eq.EOM_2holes_5}
\end{align}
where $y = |x| - a^{(n)}$, and $f(y) = \phi^{(n)}(k,y + a^{(n)})$. Hence, $y \geq -a^{(n)}$ is required here. The solutions to Eq. \eqref{eq.EOM_2holes_5} are the Airy functions ${\rm Ai}, {\rm Bi}$, such that $f(y) = A \cdot {\rm Ai}(y) + B\cdot {\rm Bi}(y)$. Normalization of the wave function then dictates that $B = 0$, i.e. $\phi^{(n)}(k,|x|) = A \cdot {\rm Ai}(|x| - a)$. Since the potential is even in $x$, we may choose eigenfunctions that are either even or odd. For even functions, the derivate of $\phi$ at $x = 0$ is
\begin{align}
\frac{d\phi^{(n)}(k,x)}{dx}\Big|_{x = 0^{\pm}} = \pm A\frac{d{\rm Ai}(y)}{dy}\Big|_{y = -a^{(n)}}.
\label{eq.EOM_2holes_5}
\end{align}
Since the potential is continuous everywhere, so must the derivative be. This, in particular, holds at $x = 0$, and, therefore, $-a^{(n)}$ must be a zero of the derivative of the Airy function, ${\rm Ai}'(-a^{(n)}) = 0$. This defines one set of eigenfunctions with the lowest eigenvalues given by $a_e = 1.01879.., 3.24819.., 4.82010.., \dots$. \\

For odd functions, we need $\phi^{(n)}(k,x) = A \cdot {\rm sgn}(x) {\rm Ai}(|x| - a^{(n)})$ to vanish at $x = 0$. Hence, $-a^{(n)}$ must be a zero of the Airy function itself, ${\rm Ai}(-a^{(n)}) = 0$. This defines another set of eigenvalues: $a_o = 2.33811.., 4.08795.., 5.52056.., \dots$. As one can expect, we get an alternating pattern of even and odd eigenstates. The asymptotic energies are, hereby, given by
\begin{equation}
E_{2\perp}^{(n)}(k) = -4t \cos\left(\frac{k}{2}\right)\left[1 - \frac{a^{(n)}}{2}\lambda^2(k)\right] + \frac{J_\perp}{2},
\label{eq.E_2_asymp}
\end{equation}
with $\lambda(k) = [J_\perp/(4t\cos(k/2))]^{1/3}$, and where $a^{(2m)} = a_e^{(m)}$ and $a^{(2m + 1)} = a_e^{(m)}$ for even $n = 2m$ and odd $n = 2m + 1$ eigenstates, respectively. The asymptotic eigenstates for holes on separate legs are, hereby,
\begin{equation}
\psi^{(n)}(k,d) = A \cdot ({\rm sgn}(d))^{n} \sqrt{\lambda(k)} \cdot {\rm Ai}(\lambda(k)|d| - a^{(n)}),
\label{eq.psi_n_k_d_asymp}
\end{equation}
with normalization constants $A_n$. The full derivation, here, carries over to two holes on the same leg. However, in this case the hard-core constraint of the holes mean that the wave function must vanish at $d = 0$. Consequently, only the odd asymptotic wave functions, $\psi^{(2n+1)}(k,d)$, from above are allowed in this case, and hence
\begin{equation}
E_{2\parallel}^{(n)}(k) = -4t \cos\left(\frac{k}{2}\right)\left[1 - \frac{a^{(2n+1)}}{2}\lambda^2(k)\right] + \frac{J_\perp}{2},
\label{eq.E_2_asymp_same_leg}
\end{equation}
The lowest two-hole eigenstates are at vanishing total momentum, $k = 0$, and for $a^{(0)} = 1.01879..$ and $a^{(1)} = 2.33811..$ for holes on separate legs and the same legs, respectively,
\begin{align}
E_{2\perp}^{(0)}(0) &= -4t\left[1 - \frac{a^{(0)}}{2}\left(\frac{J_\perp}{4t}\right)^{2/3}\right] - \frac{J_\perp}{2}, \nn\\
E_{2\parallel}^{(0)}(0) &= -4t\left[1 - \frac{a^{(1)}}{2}\left(\frac{J_\perp}{4t}\right)^{2/3}\right] - \frac{J_\perp}{2}.
\label{eq.E_2_asymp_zero_momentum}
\end{align}
We note that for a fixed $J_\perp / t$, Eqs. \eqref{eq.E_2_asymp} and \eqref{eq.psi_n_k_d_asymp} will break down as one approaches $k = \pm \pi$. Finally, a very similar calculation shows that asymptotically, the single-hole energy is
\begin{equation}
E_{1}^{(n)} = -2t\left[1 - \frac{a^{(n)}}{2}\left(\frac{J_\perp}{2t}\right)^{2/3}\right] + \frac{J_\parallel}{2}.
\label{eq.E_1_asymp}
\end{equation}
Equations \eqref{eq.E_1_asymp} and \eqref{eq.E_2_asymp_zero_momentum} give the asymptotic binding energy in Eq. \eqref{eq.binding_energy_asymptotic}.

\section{Quantum walks of two holes} \label{app.quantum_walks}
In this appendix, we derive the probability distributions in Eq. \eqref{eq.probability_distribution_distance_free} describing the distance between two non-interacting particles performing quantum walks either in the separate-legs or same-leg configuration.

The hopping Hamiltonian for identical particles may simply be written as
\begin{align}
\Ham_t = - t\sum_{j,\mu} \left[\hat{c}^\dagger_{j,\mu} \hat{c}_{j + 1,\mu} + \hat{c}^\dagger_{j + 1,\mu} \hat{c}_{j,\mu}\right].
\label{eq.H_t}
\end{align}
To easily enforce the hard-core constraint in the case of particles on the same leg, we use fermionic commutation relations $\{\hat{c}_{j,\mu}, \hat{c}^\dagger_{l,\nu}\} = \delta_{j,l}\delta_{\mu,\nu}$. The Hamiltonian is diagonalized by Fourier transforming to crystal momentum states,  $\hat{c}^\dagger_{j,\mu}  = \sum_k \te^{-ikj} \hat{c}_{k,\mu} / \sqrt{N}$,
\begin{align}
\Ham_t = \sum_{k,\mu} \varepsilon_k \hat{c}^\dagger_{k,\mu} \hat{c}_{k,\mu},
\label{eq.H_t}
\end{align}
with $\varepsilon_k = -2t\cos(k)$. From the initial states $\ket{\Psi_\perp(\tau = 0)} = \hat{c}^\dagger_{0,1}\hat{c}^\dagger_{0,2} \ket{0}, \ket{\Psi_\parallel(\tau = 0)} = \hat{c}^\dagger_{0,1}\hat{c}^\dagger_{1,1} \ket{0}$, we find the non-equilibrium states 
\begin{align}
\ket{\Psi_\perp(\tau)} &= \frac{1}{N}\sum_{k,q} \te^{-i(\varepsilon_k + \varepsilon_q)\tau}  \hat{c}^\dagger_{k,1}\hat{c}^\dagger_{q,2} \ket{0} \nn \\
\ket{\Psi_\parallel(\tau)} &= \frac{1}{N}\sum_{k,q} \te^{-iq} \te^{-i(\varepsilon_k + \varepsilon_q)\tau}  \hat{c}^\dagger_{k,1}\hat{c}^\dagger_{q,1} \ket{0}.
\end{align}
So far, there is hardly any difference between the two cases. This, however, appears when we compute the amplitude for seeing the particles at positions $j,j+d$
\begin{widetext}
\begin{align}
A_\perp(j,d,\tau) &= \bra{0}\hat{c}_{j,1}\hat{c}_{j+d,2}\ket{\Psi_\perp(\tau)} = \frac{1}{N^2}\!\sum_{\substack{k_1,k_2 \\ q_1,q_2}} \!\te^{i(k_2 + q_2)j + q_2d}  \te^{-i(\varepsilon_{k1} + \varepsilon_{q1})\tau} \!\bra{0} \hat{c}_{q_2,2}\hat{c}_{k_2,1}\hat{c}^\dagger_{k_1,1}\hat{c}^\dagger_{q_1,2} \ket{0},\nn \\
A_\parallel(j,d,\tau) &= \bra{0}\hat{c}_{j,1}\hat{c}_{j+d,1}\ket{\Psi_\parallel(\tau)} = \frac{1}{N^2}\!\sum_{\substack{k_1,k_2 \\ q_1,q_2}} \!\te^{i(k_2 + q_2)j + q_2d}\te^{-iq_1} \te^{-i(\varepsilon_{k1} + \varepsilon_{q1})\tau}\! \bra{0} \hat{c}_{q_2,1}\hat{c}_{k_2,1}\hat{c}^\dagger_{k_1,1}\hat{c}^\dagger_{q_1,1} \ket{0},
\end{align}
\end{widetext}
because the particles on separate legs only have a single nonzero matrix element $\bra{0} \hat{c}_{q_2,2}\hat{c}_{k_2,1}\hat{c}^\dagger_{k_1,1}\hat{c}^\dagger_{q_1,2} \ket{0} = \delta_{k_1,k_2}\delta_{q_1,q_2}$, whereas particles on the same leg also feature an exchange term $\bra{0} \hat{c}_{q_2,2}\hat{c}_{k_2,1}\hat{c}^\dagger_{k_1,1}\hat{c}^\dagger_{q_1,2} \ket{0} = \delta_{k_1,k_2}\delta_{q_1,q_2} - \delta_{k_1,q_2}\delta_{q_1,k_2}$. As a result, the amplitudes simplify to
\begin{align}
A_\perp(j,d,\tau) &= \frac{1}{N^2}\!\sum_{k,q} \!\te^{iqd} \te^{i(k + q)j} \te^{-i(\varepsilon_{k} + \varepsilon_{q})\tau}, \nn \\
A_\parallel(j,d,\tau) &= \frac{1}{N^2}\!\sum_{k,q} \!\big[ \te^{iqd}  -  \te^{ikd} \big]\te^{i(k + q)j} \te^{-iq}\te^{-i(\varepsilon_{k} + \varepsilon_{q})\tau}.
\label{eq.amplitudes_quantum_walk}
\end{align}
From here, we may, then, calculate the probabilities to find the holes a distance $d$ apart. Since we are not interested in the absolute position of the holes, $j$, we get
\begin{align}
&P^{\rm q.w.}_\perp(d,\tau) = \sum_j |A_\perp(j,d,\tau)|^2 \nn \\
&= \frac{1}{N^4}\!\sum_{\substack{j,k_1,k_2 \\ q_1,q_2}} \!\te^{i(q_1 - q_2)d} \te^{i(k_1 - k_2 + q_1 - q_2)j} \te^{-i(\varepsilon_{k1} + \varepsilon_{q1} - \varepsilon_{k2} - \varepsilon_{q2})\tau} \nn \\
&= \frac{1}{N^3}\!\sum_{k,q,p} \!\te^{ipd} \te^{i(\varepsilon_{k+p} + \varepsilon_{q-p} - \varepsilon_{k} - \varepsilon_{q})\tau} \nn \\
&= \frac{1}{N^3}\!\sum_{k,q,p} \!\te^{ipd} \te^{i(\varepsilon_{k+p} - \varepsilon_{k})\tau}  \te^{-i(\varepsilon_{q+p} - \varepsilon_{q})\tau}\nn \\
&= \frac{1}{N}\!\sum_{p} \!\te^{ipd} \Big|\frac{1}{N}\sum_k\te^{i(\varepsilon_{k}-\varepsilon_{k+p})\tau}\Big|^2 
\end{align}
Combining the summands at $p$ and $-p$ then results in top line of Eq. \eqref{eq.probability_distribution_distance_free} describing the probability distribution for the distance between the holes on separate legs. A completely analogous calculation derives the bottom line of Eq. \eqref{eq.probability_distribution_distance_free} from the bottom line of \eqref{eq.amplitudes_quantum_walk} for two holes on the same leg.

\bibliographystyle{apsrev4-2}
\bibliography{ref_magnetic_polaron}

\end{document}